%% file: main.tex
\documentclass[11pt,a4paper]{article}
\usepackage{jinstpub}

\usepackage{multirow,booktabs}
\usepackage{subcaption}

\newcommand{\NN}{\ensuremath{\mathrm{N}_{2}}}
\newcommand{\HH}{\ensuremath{\mathrm{H}_{2}}}

\newcommand{\CN}{\mbox{CEDAR-N}}
\newcommand{\CW}{\mbox{CEDAR-W}}
\newcommand{\CH}{\mbox{CEDAR-H}}

\title{Development of a new CEDAR for kaon identification at the NA62 experiment at CERN}

\author{The NA62 collaboration}

\emailAdd{na62eb@cern.ch}

\abstract{The NA62 experiment at CERN utilises a differential Cherenkov counter with achromatic ring focus (CEDAR) for tagging kaons within an unseparated monochromatic beam of charged hadrons. The \CH\ detector was developed to minimise the amount of material in the path of the beam by using hydrogen gas as the radiator medium.
The detector was shown to satisfy the kaon tagging requirements in a test-beam before installation and commissioning at the experiment.
The \CH\ performance was measured using NA62 data collected in 2023.}


\arxivnumber{2312.17188}

\begin{document}

\maketitle








\section{Introduction} 

\noindent The NA62 experiment at CERN~\cite{NA62_Det} is designed to measure the branching fraction of the \mbox{$K^{+}\to\pi^{+}\nu\bar{\nu}$ } decay, predicted to be $(8.4\pm 1.0)\times10^{-11}$~\cite{PNN2015}, as a stringent test of the Standard Model.
First results based on data collected in 2016--2018 have been published~\cite{PNN2021}.

\begin{figure}[ht]
\centering
\includegraphics[width=1.0\textwidth]{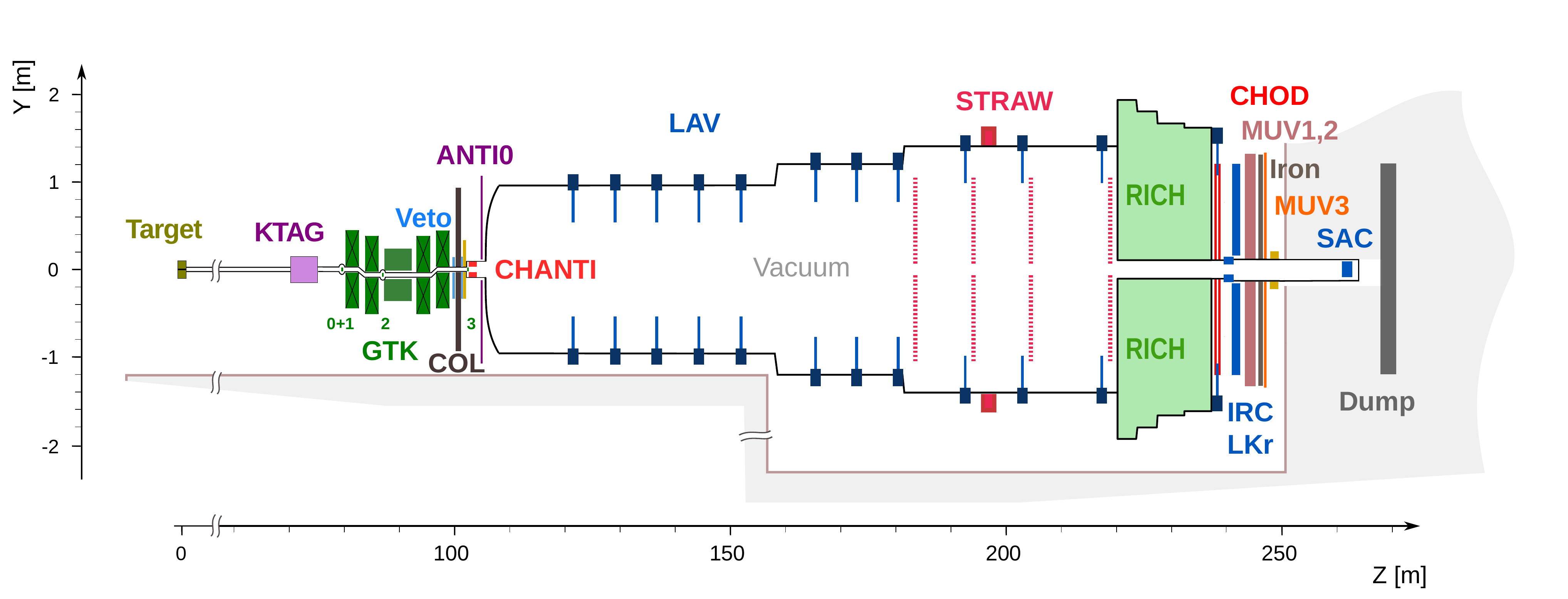}
\caption{Schematic side view of the NA62 beamline and detector in 2021.}
\label{fig:na62-sketch}
\end{figure}

The layout of the NA62 beamline and detector is shown schematically in figure~\ref{fig:na62-sketch}.
An unseparated secondary hadron beam is produced by directing 400\,GeV/$c$ protons extracted from the CERN SPS onto a beryllium target in spills of 4.8\,s duration.
The nominal particle rate in the beam is 600\,MHz, comprising $\pi^{+}$ (70\%), protons (23\%) and $K^{+}$ (6\%).
The central beam momentum is 75\,GeV/$c$, with a spread of 1\% (rms).
The beam travels mostly in vacuum from the target, through a fiducial volume that extends between 105\,m and 180\,m from the target, and ends in a beam dump. 
The experiment is equipped with a beam spectrometer (GTK) composed of four silicon-pixel detector stations, with the most upstream station located 80\,m from the target and the most downstream station (GTK3) located 102\,m from the target; a STRAW spectrometer located downstream of the fiducial volume, between 180 and 220\,m from the target; and hodoscopes (CHOD) located 238\,m from the target.
The other principal subdetectors are two Cherenkov counters (KTAG, RICH), a hermetic photon veto (LAV, LKr, IRC, SAC), a hadronic calorimeter (MUV1,2) and a muon system (MUV3).
Data from the detectors are collected via a two-stage trigger system, with the first stage implemented in hardware and the latter implemented in software.

Charged kaons are tagged to suppress backgrounds from interactions of beam pions with material on the beamline.
Kaon tagging is achieved via the KTAG -- a differential Cherenkov counter with achromatic ring focus (CEDAR)~\cite{BOVET} connected to a bespoke photon-detection system~\cite{KTAG}.
The $K^{+}$ must be identified with efficiency above 95\%, and the kaon--pion separation must be better than $10^{4}$, 
meaning that for each misidentified pion there are more than $10^{4}$ correctly identified kaons. 
The time resolution must be better than 100\,ps to provide a precise time reference for event reconstruction and selection.

The CEDAR was originally developed at CERN with two variants -- North (\CN) and West (\CW) -- adapted for different beam momenta~\cite{BOVET}.
At NA62, the CEDAR extends from 70\,m to 75\,m downstream of the target.
The CEDAR utilises a gaseous radiator to produce Cherenkov light, and each end of the gas vessel connects to a vacuum beam pipe.
Gas in the vessel is isolated from the vacuum in the beam pipe by two aluminium windows of thickness $150\,\mu$m and $200\,\mu$m.
Both the gas and the aluminium windows contribute to the material in the path of the beam.

From the start of NA62 data-taking in 2016, the KTAG used a \CW\ filled with nitrogen gas (\NN) at 1.71\,bar (in this paper, pressure values refer to absolute pressure).
In this configuration, the CEDAR introduces \mbox{$39\times10^{-3}\,X_{0}$} of material in the path of the beam, where $X_{0}$ is one radiation length.
This comprises \mbox{$35\times 10^{-3}\,X_{0}$} from the \NN\ gas and $3.9\times 10^{-3}\,X_{0}$ from the aluminium windows.
Multiple scattering of the beam particles introduces an angular divergence of $32\,\mu$rad (rms) in both the horizontal and vertical planes, which adds quadratically to the nominal $70\,\mu$rad beam divergence at the CEDAR position.

Filling the CEDAR with \HH\ at 3.85\,bar, as required to achieve a similar Cherenkov angle to \NN\ at 1.71\,bar, reduces the material in the path of the beam to $7.3\times 10^{-3}\,X_{0}$. 
This comprises $3.4\times10^{-3}\,X_{0}$ from the \HH\ gas and $3.9\times 10^{-3}\,X_{0}$ from the aluminium windows.
In this case, multiple scattering introduces an angular divergence of $13\,\mu$rad (rms) in both the horizontal and vertical planes.

\begin{figure}[t]
    \centering
    \begin{subfigure}[b]{0.49\textwidth}
      \includegraphics[width=\textwidth]{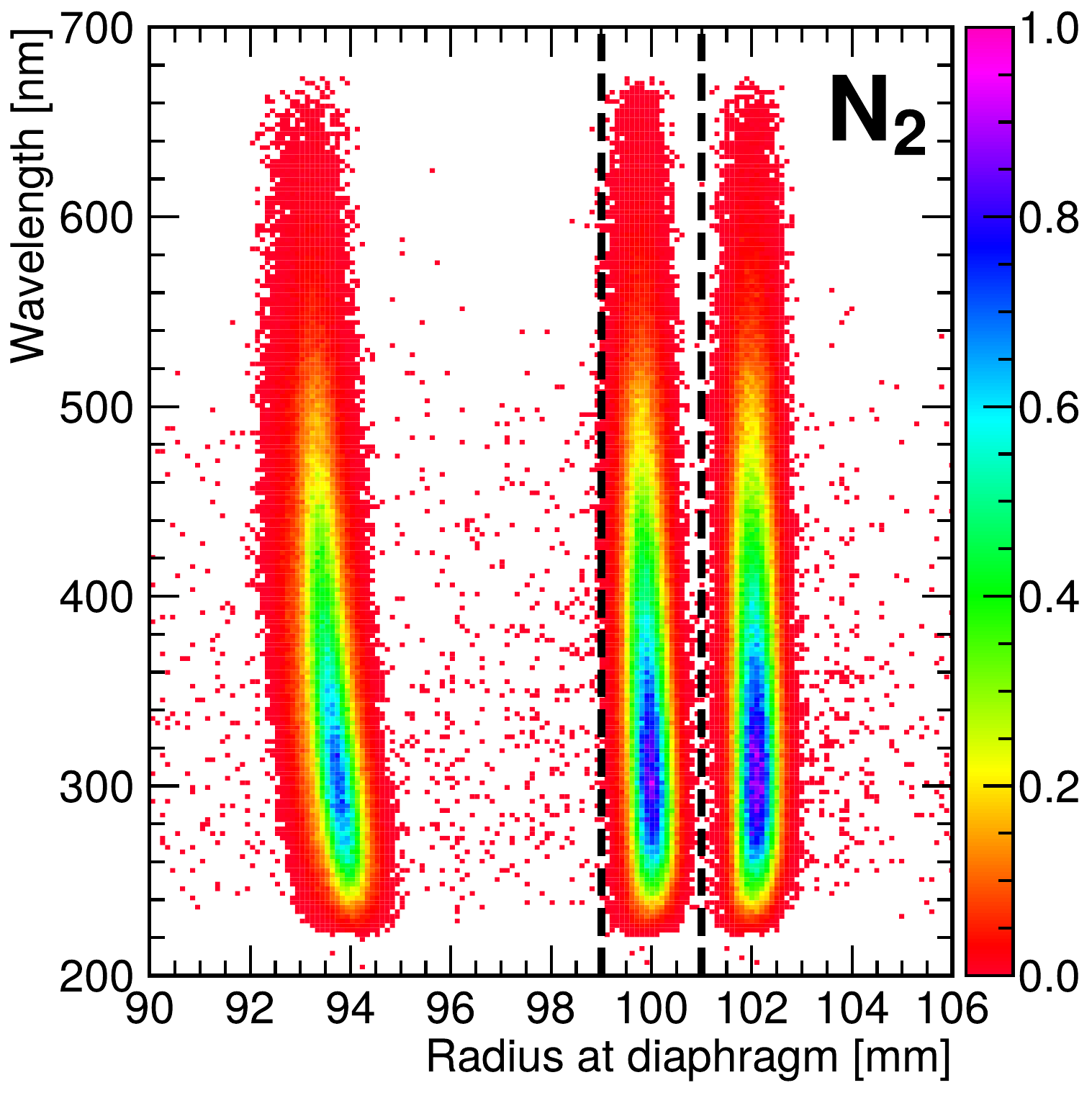}%
      \label{fig:waveVsRadiusCEDARW:left}
    \end{subfigure}
    \begin{subfigure}[b]{0.49\textwidth}
      \includegraphics[width=\textwidth]{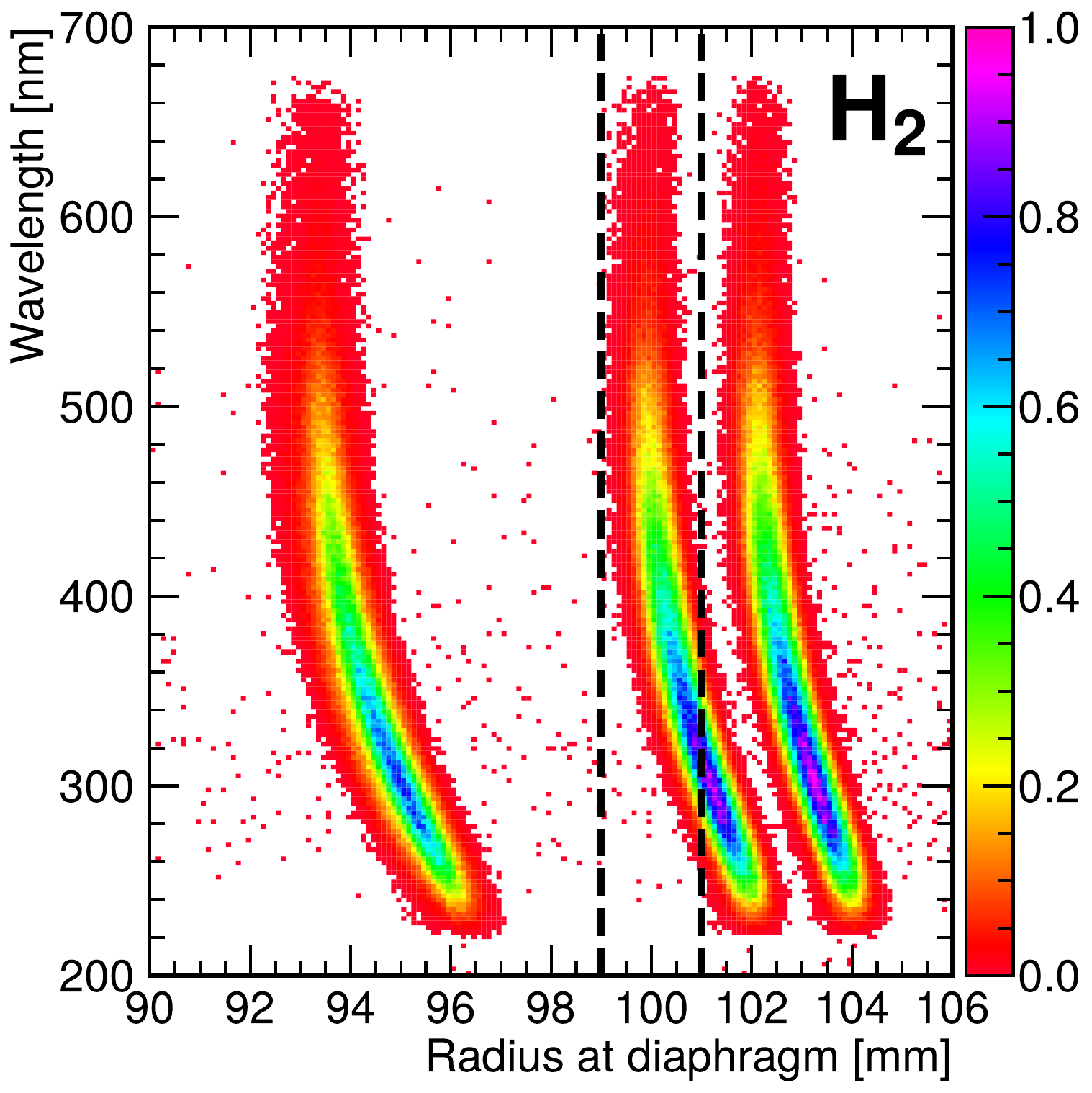}%
      \label{fig:waveVsRadiusCEDARW:right}
    \end{subfigure}
    \caption{Simulated radial position of Cherenkov photons at the location of the diaphragm in \CW\ for proton (lowest radius), kaon and pion (highest radius) beam particles as a function of wavelength with \NN\ at 1.71\,bar (left panel)
    and \HH\ at 3.67\,bar (right panel). 
    The quantum efficiency of the KTAG photomultiplier tubes and the transmittance of the glass filters are imposed. Dashed lines show the extent of a 2.0\,mm diaphragm aperture centred at 100\,mm.}
    \label{fig:waveVsRadiusCEDARW}%
\end{figure}

Considering the \mbox{$20\times10^{-3}\,X_{0}$} contribution of the four GTK stations, replacing the \NN\ with \HH\ reduces the total material in the path of the beam from \mbox{$59\times10^{-3}\,X_{0}$} to \mbox{$27\times10^{-3}\,X_{0}$}.
A full simulation of NA62 implemented with the GEANT4 toolkit~\cite{Geant4} shows that the fraction of beam particles interacting inelastically with material upstream of the fiducial volume decreases from 2.1\% to 0.9\%, in agreement with the above expectation.
The simulation also indicates a similar reduction in elastic scattering.

\textcolor{black}{The reduction of scattering in the CEDAR filled with \HH\ improves kaon transmission and reduces signal rates in the downstream detectors, and leads to a more efficient selection of  \mbox{$K^{+}\to\pi^{+}\nu\bar{\nu}$ } decays. 
Furthermore, }it is expected to improve the performance of the hardware trigger designed to collect $K^{+}\to\pi^{+}\nu\bar{\nu}$ decays.
In 40\% of the events collected via this trigger during 2022, when using the \CW\ filled with \NN, the STRAW track closest to the trigger time is consistent with an elastically-scattered beam particle originating upstream of GTK3.
This suggests that beam particles elastically scattering in \CW\ are responsible for a substantial fraction of the hardware trigger rate.
The simulation shows that a CEDAR filled with \HH\ rather than \NN\ reduces the flux of elastically-scattered beam particles passing through detectors downstream of the fiducial volume by 40\%, which would reduce the rate of the hardware trigger used to collect $K^{+}\to\pi^{+}\nu\bar{\nu}$ decays by 15\%.
The lower trigger rate will improve the stability of the data acquisition, and will allow looser trigger conditions to be imposed without an overall increase in the trigger rate.

Using \HH\ in an existing North or West-type CEDAR does not lead to satisfactory kaon tagging performance, as the chromatic dispersion in \HH\ is not corrected by the optical system, and the kaon and pion rings overlap due to their large widths.
For the \CW\ filled with \HH, a pressure of 3.67\,bar optimises the kaon--pion separation.
However, 40\% of the Cherenkov light is lost as it falls outside of a 2.0\,mm diaphragm aperture centred at 100\,mm (figure~\ref{fig:waveVsRadiusCEDARW}) and the detector is unable to satisfy the kaon tagging requirements.
As no suitable CEDAR existed, a new detector named \CH\ was developed by refitting a North-type CEDAR with optics designed specifically for use with \HH. 

\section{CEDAR description}

\begin{figure}[t]
\centering
\includegraphics[width=1.00\textwidth]{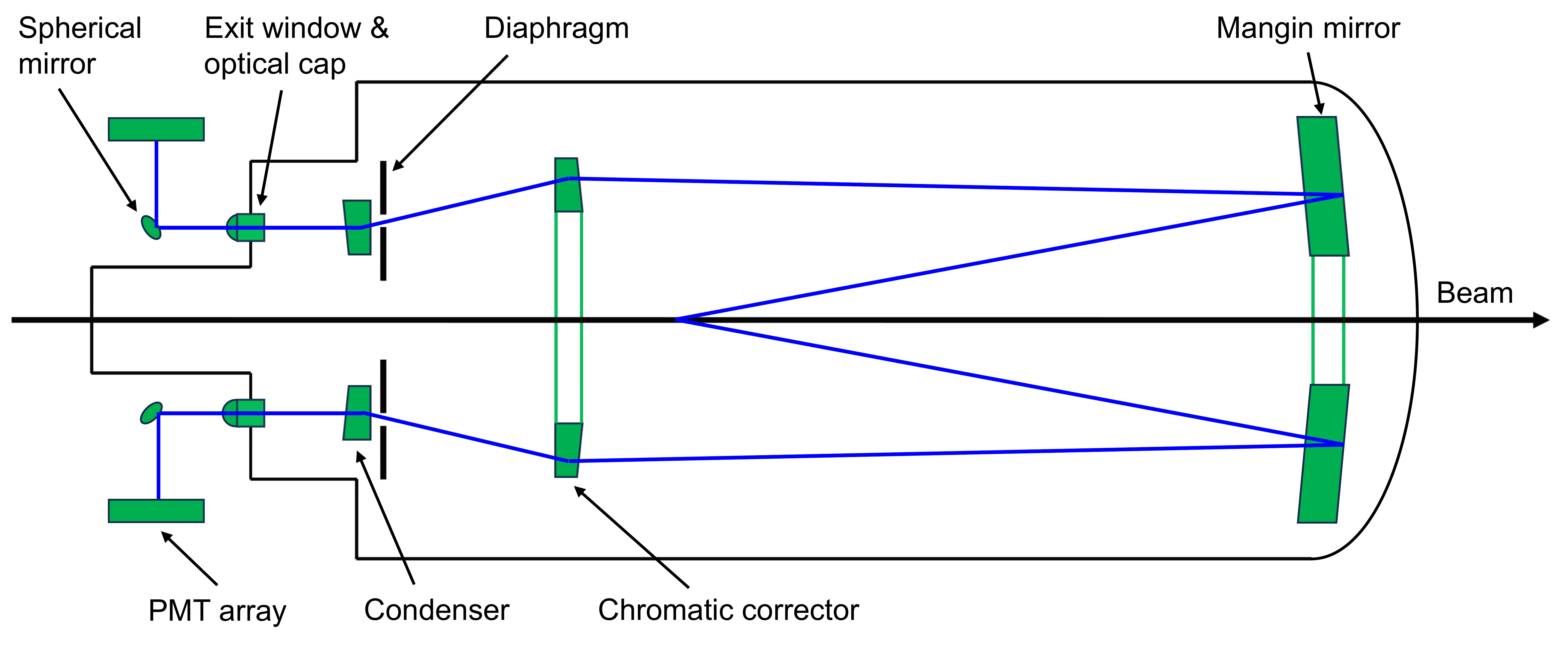}
\caption{Sketch of the KTAG optical system (not to scale), highlighting the optical elements (green areas) and the path of Cherenkov light (blue lines).}
\label{fig:cedarsketch}
\end{figure}

Each CEDAR comprises a thermally-insulated gas vessel built in two parts: a 4.5\,m long cylinder of 53\,cm inner diameter and a vessel cap either 28\,cm (\CN) or 34\,cm (\CW) long attached to the upstream end.
The hadron beam passes along the longitudinal axis of the CEDAR.

The CEDAR optical system is contained within the gas vessel, with the optical axis aligned to the longitudinal axis of the CEDAR (figure~\ref{fig:cedarsketch}).
The light produced in the gas radiator is focused to a ring of mean radius 100\,mm at the position of a diaphragm, achieved by the combination of a Mangin mirror~\cite{MANGIN} and a chromatic corrector lens.
The Mangin mirror is designed to reduce geometrical aberrations; it comprises a lens with two spherical surfaces with the downstream surface cemented onto a concave mirror.
The chromatic corrector is a plano-convex lens whose function is to compensate chromatic dispersion in the gas and lenses.
Both the Mangin mirror and chromatic corrector have a central hole for the hadron beam to pass through.

The diaphragm is an annular aperture centred at 100\,mm, whose width can be varied from zero to 20\,mm. 
Cherenkov light that passes through the aperture traverses condenser lenses that direct the light out of the gas vessel via eight quartz exit windows.
The exit windows are equally spaced in azimuth, encircling the beam pipe on the upstream end of the CEDAR.
Filters made of glass are glued to the outside of the exit windows to attenuate light with wavelength below $240$\,nm.

In the original CEDAR design, light exiting the CEDAR is detected by eight ET 9820QB photomultiplier tubes (PMTs), with one PMT positioned on the outside of each exit window.
These PMTs cannot sustain the particle rate at NA62, so a new photon-detection system was developed~\cite{KTAG}.
In the NA62 setup (figure~\ref{fig:cedarsketch}) the light travels through optical caps -- a set of lenses attached to the outer frames of the exit windows -- and passes into the KTAG photon-detection system, which is housed in an enclosure. 
The light reaches eight spherical mirrors and is reflected through 90 degrees to eight KTAG sectors.
Each sector is a PMT array $16 \times 20\,\mathrm{cm}^{2}$ in size equipped with 48 Hamamatsu PMTs, 32 of type R9880U-110 and 16 of type R7400U-03, which are sensitive to light with wavelength above $230$\,nm.
Spherical mirrors are used to broaden the light spot and reduce the intensity of photons on the PMTs.

The radius of the Cherenkov rings produced by beam particles passing through the CEDAR depends on the density of gas inside the vessel; increasing the gas pressure increases the refractive index, hence yields larger Cherenkov rings. As such,
the gas pressure can be set to obtain a kaon ring of radius 100\,mm, matching the central radius of the diaphragm aperture.
For gas at a temperature of 293\,K and beam particles with momenta of 75\,GeV/$c$, the appropriate gas pressure is 1.71\,bar for \CW\ and 3.85\,bar for \CH, with a change of 0.34\% in these values for each 1\,K of temperature difference.
Once the operating pressure has been established, the CEDAR is designed to operate with a fixed amount of gas and is therefore insensitive to gradual changes in temperature of the environment.
The CEDAR is nevertheless enclosed by thermal insulation to avoid any rapid change in ambient temperature causing differences in the refractive index of the gas at different places within the gas vessel~\cite{BOVET}.

With the pressure set to obtain a kaon ring of radius 100\,mm, the pion and proton rings have radii 102\,mm and 94\,mm, respectively.
Any diaphragm aperture less than 2.5\,mm is therefore sufficient to separate the kaon ring from the pion and proton rings.
However, beam particles travelling at an angle of $100\,\mu$rad with respect to the CEDAR optical axis will produce a Cherenkov ring shifted by 0.5\,mm at the diaphragm.
Thus, to ensure that the Cherenkov rings are centred on the diaphragm, the beam divergence must be below $100\,\mu$rad, and the CEDAR must be parallel to the beam within $100\,\mu$rad.
The CEDAR is fixed at the upstream end, and can be rotated parallel to the beam via motorised stages situated at its downstream end that can move by $\pm$5\,mm horizontally and vertically.
Positional offsets of the beam have no effect due to the spherical surface of the Mangin mirror.

\section{\CH\ development}
\label{sec3}

A \CN\ already available at CERN provided the structure of \CH.
No changes to the diaphragm or the optical support structure were envisaged due to their complexity, fixing the positions and outer radii of the Mangin mirror and chromatic-corrector lens, and the position and central radius of the diaphragm.
Given these constraints, initial \CH\ optical parameters were determined analytically via a ray tracing procedure that sampled the wavelength spectrum and the point of origin of the Cherenkov light.
The problem was reduced to two dimensions by exploiting the cylindrical symmetry of the CEDAR.
The radii of curvature of the corrector lens and the two surfaces of the Mangin mirror were varied iteratively to achieve a ring of 100\,mm radius at the diaphragm with minimal RMS spread. 
As the relative effects of spherical aberration and chromatic dispersion vary with the gas pressure, the computation was repeated between 3.7 and 4.1\,bar in steps of 0.1\,bar at a temperature of 293\,K, aiming to obtain the best solution at the lowest pressure.

A complete simulation of the KTAG was then employed to refine the analytical solutions using an iterative procedure to minimise the width of the Cherenkov ring.
The simulation was then used to identify modifications to the optical system that maximise the amount of light reaching the PMT arrays: the size of the central hole in the Mangin mirror was reduced with respect to the original CEDAR design, making the reflective surface larger, and the radius of curvature of the spherical mirrors was increased so that the distribution of light matched the size and shape of the PMT arrays.
The optimum light propagation through the KTAG was achieved with condenser lenses taken from a \CW\ rather than a \CN.
It was not necessary to change the optical caps.

\begin{figure}[t]
    \centering
    \begin{subfigure}[b]{0.49\textwidth}
      \includegraphics[width=\textwidth]{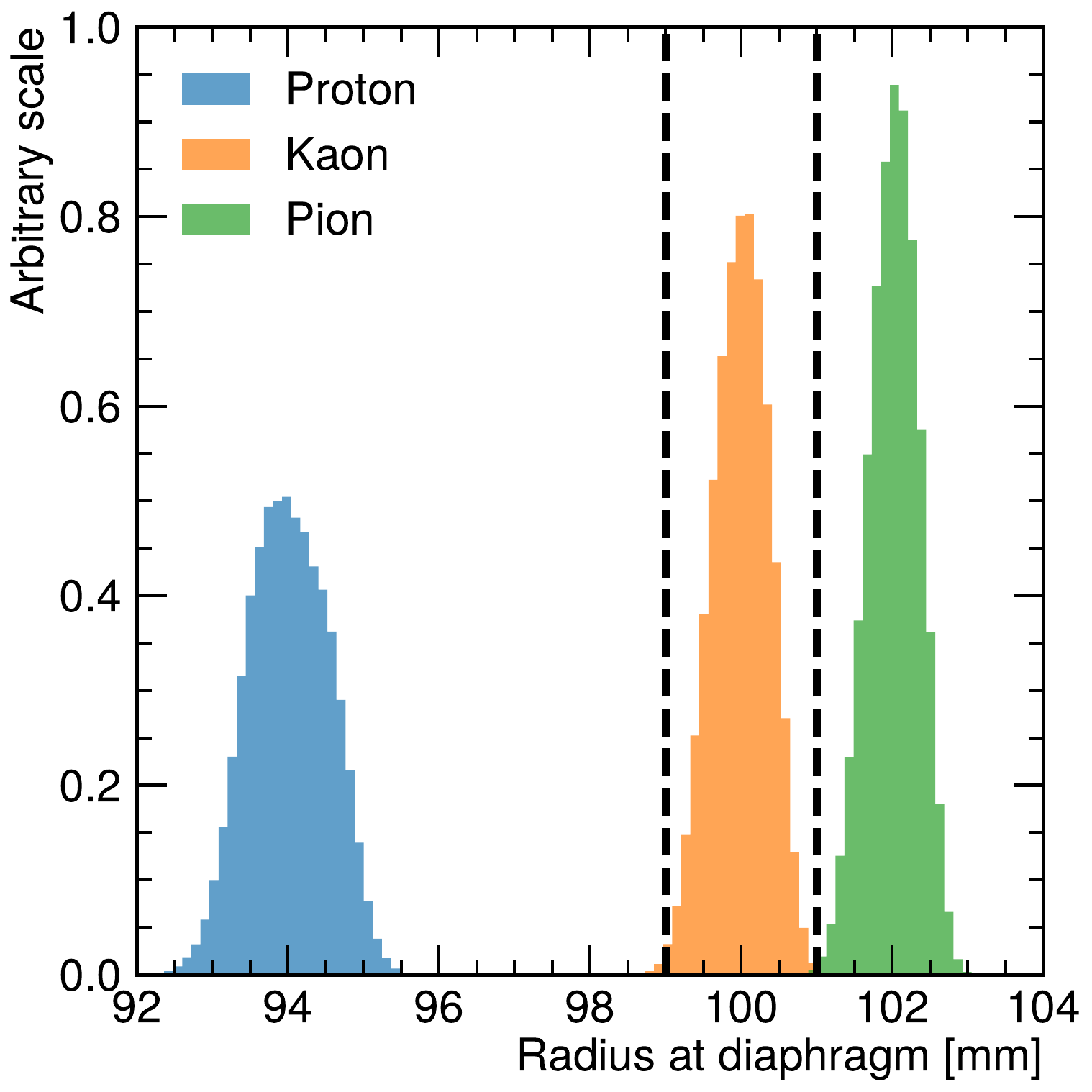}%
      \label{fig:waveVsRadius:left}
    \end{subfigure}
    \begin{subfigure}[b]{0.49\textwidth}
      \includegraphics[width=\textwidth]{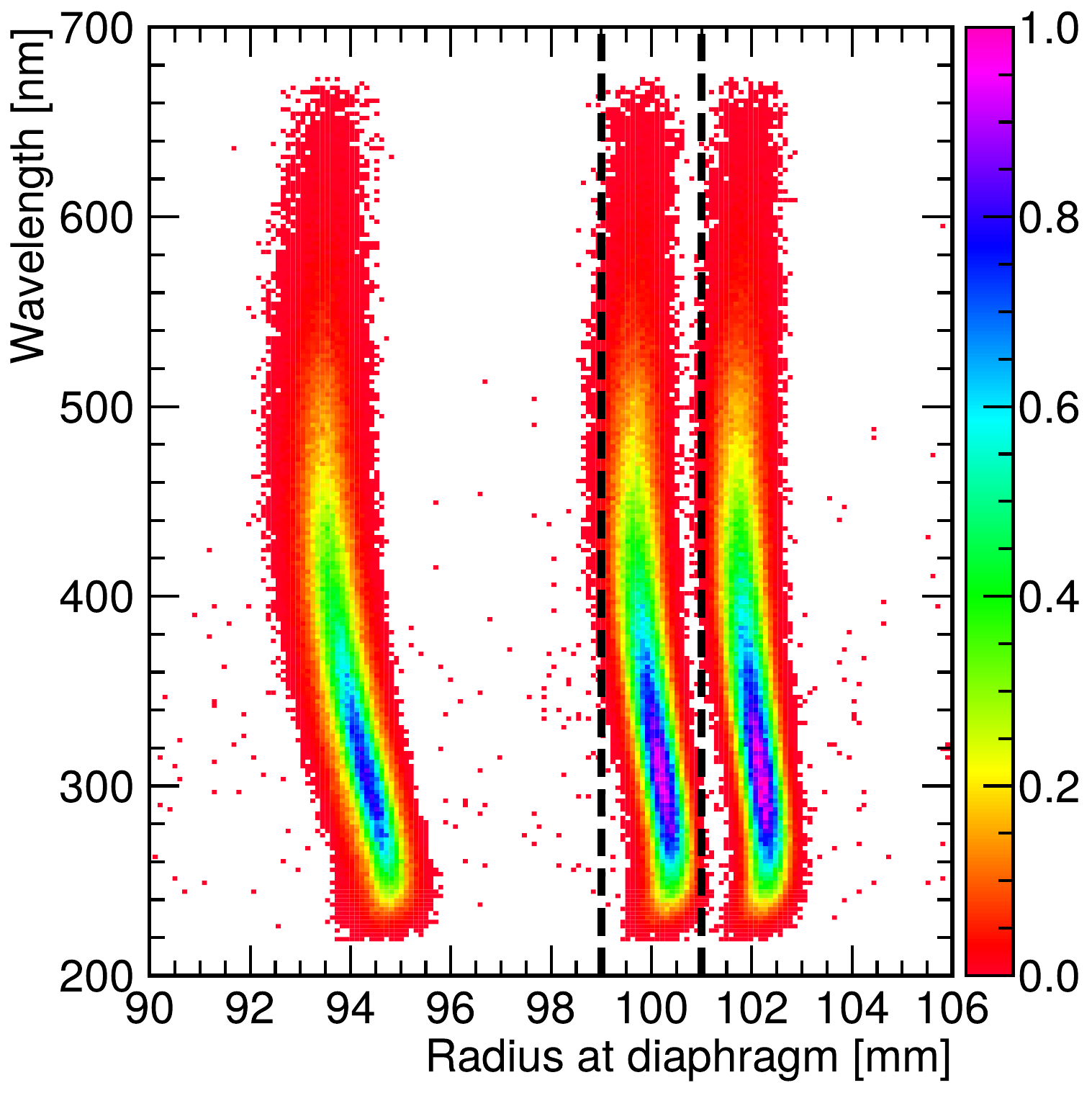}%
      \label{fig:waveVsRadius:right}
    \end{subfigure}
    \caption{Simulated radial position at the diaphragm of Cherenkov photons for proton, kaon and pion beam particles with \CH\ at 3.8\,bar (left panel), and as a function of wavelength (right panel). The quantum efficiency of the KTAG photomultipliers and the transmittance of the glass filters are imposed. The three distributions are normalised to the same integral. Dashed lines show the extent of a 2.0\,mm diaphragm aperture centred at 100\,mm.}
    \label{fig:waveVsRadius}%
\end{figure}

The results of the optimisation showed negligible difference in performance between 3.8\,bar and 4.1\,bar despite greater production of Cherenkov light at higher pressure, primarily because the light spot better matches the PMT arrays at lower pressure.
The lower pressure of 3.8\,bar is chosen, motivated by \HH\ safety considerations. 
At this pressure, the chromatic dispersion is sufficiently corrected so that the Gaussian width of the kaon and pion rings is 0.4\,mm \mbox{and the rings do not overlap}.
However, the proton ring is twice as large and distorted due to chromatic dispersion effects (figure~\ref{fig:waveVsRadius}).

The identification of a particle in the KTAG is defined by detecting coincident light in multiple sectors.
Requiring a coincidence in a larger number of sectors reduces the $K^{+}$ identification efficiency, while requiring a coincidence in a smaller number of sectors increases the pion misidentification probability.
A 5-fold sector coincidence has been the standard kaon tagging requirement since 2016~\cite{NA62_Det}, and corresponds to light detected in 5 or more sectors. 
In the \CH\ simulation, this requirement leads to $K^{+}$ identification efficiency of 99.5\% and pion misidentification probability below $10^{-4}$ for a 2\,mm diaphragm aperture.
These values exceed the kaon tagging requirements.

\begin{figure}[t]
\centering
\includegraphics[width=0.95\textwidth]{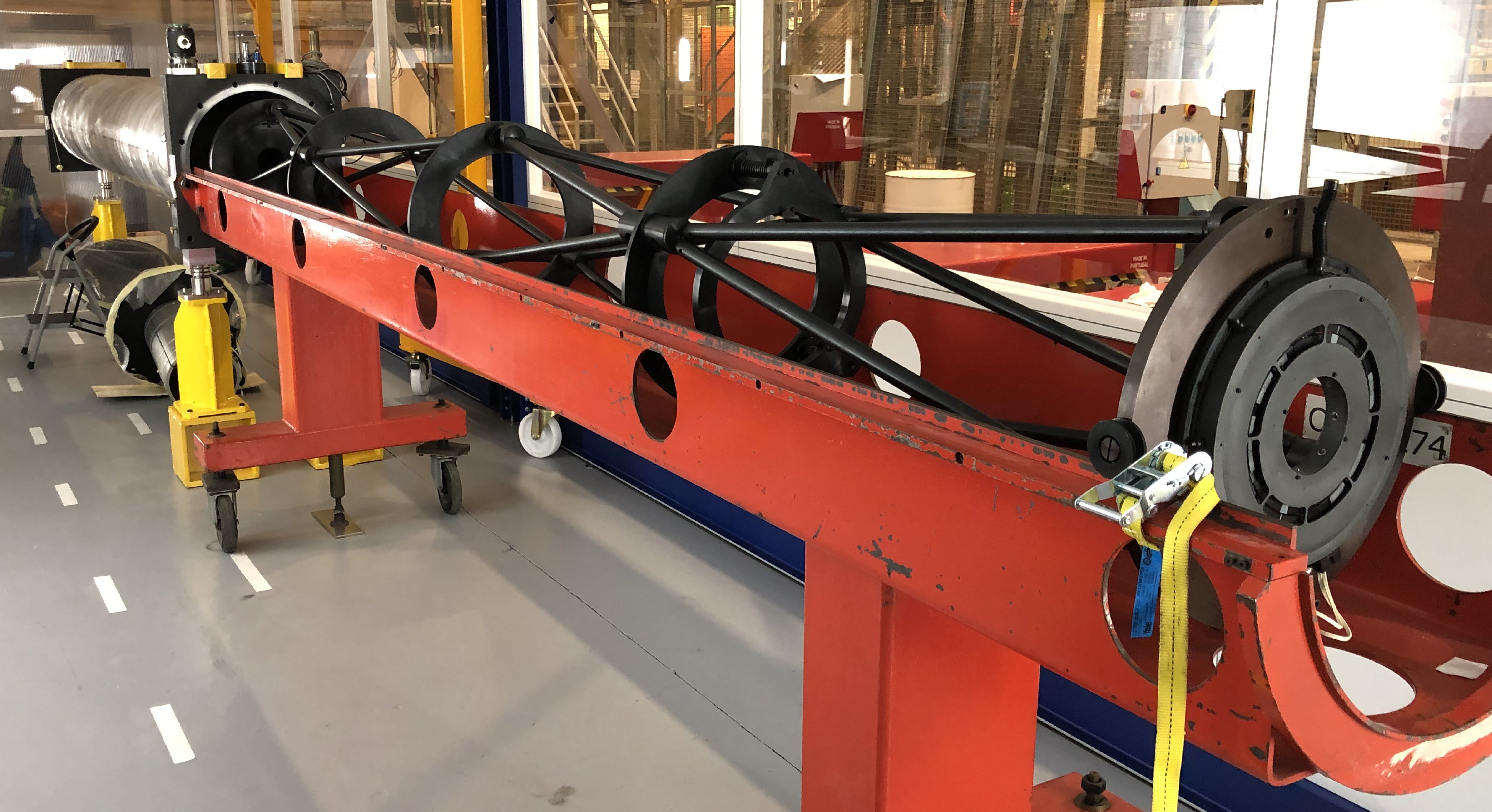}
\caption{\CH\ under construction in a clean room at CERN. The photo shows the cylindrical gas vessel (silver-coloured part on the left of the image) and the support structure of the optical system (black part) sitting on a specialised CEDAR workbench (red part). The diaphragm can be seen at the end of the support structure on the right of the image.}
\label{fig:cedarconstructionphoto}

\bigskip\bigskip\bigskip\bigskip

\includegraphics[width=0.95\textwidth]{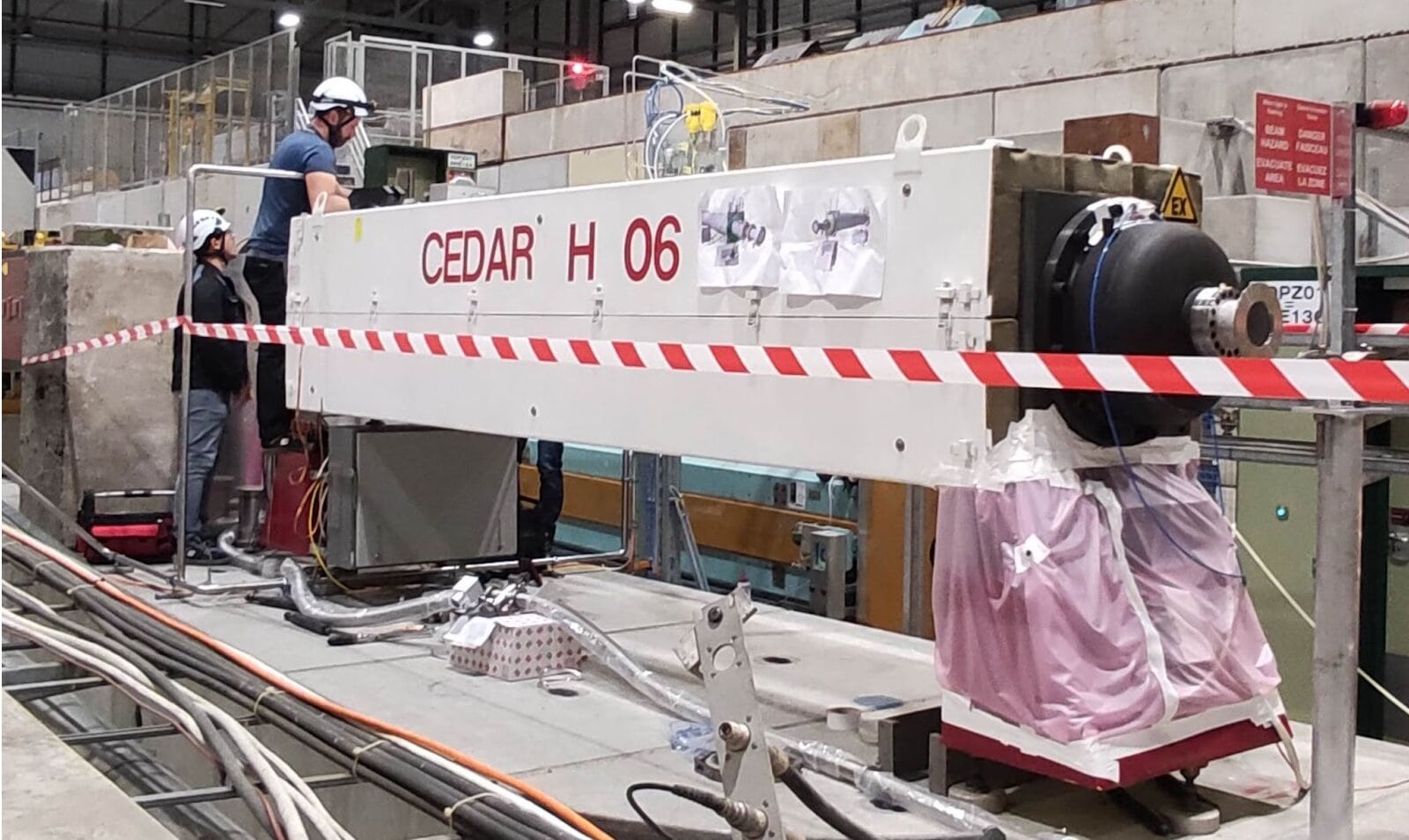}
\caption{\CH\ in preparation for the test-beam on the H6 beamline at CERN.}
\label{fig:intall_testbeam}

\end{figure}

The list of \CH\ and \CW\ mechanical and optical parameters is given in table~\ref{table:CedarParameterValues}.
The \CH\ was constructed at CERN in 2022 (figure~\ref{fig:cedarconstructionphoto}).
The Mangin mirror and chromatic corrector were fabricated from high-quality quartz blanks with sub-micron tolerances on their shape.
A laser-based procedure was used to align the optical components with microradian precision, well within the \CH\ operating tolerance.
New lenses for the spherical mirrors were coated with 90\,nm of aluminium as a reflective layer and 10\,nm of SiO$_{2}$ as a protective layer.
The spherical mirrors were aligned using a laser setup to ensure the Cherenkov light is centred on each PMT array.
Millimetre precision was achieved on the light spot positions at the PMT arrays.

\begin{table}[p]
\centering
\caption{Optical and mechanical parameters of \CW\ and \CH. All values are dimensions in millimetres unless otherwise stated. Positions along the beam axis are quoted with respect to the upstream end of the CEDAR.}
\label{table:CedarParameterValues}

\begin{tabular}{@{}llrr@{}}
\textbf{CEDAR type} &   & \textbf{\CW} & \textbf{\CH} \\ 
\bottomrule

Nominal gas type &   & \NN & \HH \\
\multicolumn{2}{@{}l}{Nominal pressure [bar]} & 1.71 & 3.80 \\
\midrule

\multirow{2}{*}{Gas vessel cylinder}       & Length      & 4500 & 4500 \\
& Inner radius & 267  & 267 \\
\midrule

\multirow{2}{*}{Gas vessel cap}      & Length       & 339 & 280 \\
& Inner radius  & 139 & 139 \\
\midrule

\multirow{5}{*}{Chromatic corrector} & Position along the beam axis& 1855 & 1902 \\
& Radius of curvature  & 1385 & 1307 \\
& Central thickness    & 20   & 20 \\
& Inner radius         & 75   & 75 \\
& Outer radius         & 160  & 160 \\
\midrule

\multirow{6}{*}{Mangin mirror}       & Position along the beam axis& 5353 & 5362 \\
& Radius of curvature: & &  \\
& ~~~~~~~ - refracting surface & 6615 & 8994 \\
& ~~~~~~~ - reflecting surface & 8610 & 9770 \\
& Central thickness                & 40   & 40 \\
& Inner radius                     & 50   & 40 \\
& Outer radius                     & 150  & 150 \\
\midrule

\multirow{2}{*}{Diaphragm}           & Position along the beam axis                 & 872 & 911 \\
& Aperture central radius & 100  & 100 \\
\midrule

\multirow{3}{*}{Condensers}          & Position along the beam axis & 832 & 871 \\
& Maximum thickness    & 10   & 10 \\
& Radius of curvature  & 300  & 300 \\
\midrule

\multirow{4}{*}{Quartz windows}      & Position along the beam axis    & 472 & 531 \\
& Thickness                 & 10  & 10 \\
& Radius                    & 22.5  & 22.5 \\
& Radial distance to window centre & 103 & 103 \\ 
\bottomrule\bottomrule

\multirow{3}{*}{Optical caps}          & Position along the beam axis & 450 & 450 \\
& Maximum thickness    & 4.24   & 4.24 \\
& Radius of curvature  & 114.62  & 114.62 \\
\midrule

\multirow{5}{*}{Spherical mirrors}   & Position along the beam axis             & 322 & 322 \\
& Radius of curvature           & 51.68 & 77.52\\
& Diameter                      & 50  & 50 \\
& Radial distance to mirror centre & 106 & 106 \\ 
\end{tabular}

\end{table}

\clearpage

\section{Test-beam at CERN}
\label{sec4}

\noindent The \CH\ performance was measured at a test-beam in October 2022 on the H6 beamline at CERN (figure~\ref{fig:intall_testbeam}).
The aims were twofold: to validate the performance of the optical components and their alignment inside the gas vessel, and to measure the $K^{+}$ identification efficiency and kaon--pion separation.

Both the NA62 and the H6 hadron beams are derived from interactions of the 400\,GeV/$c$ primary proton beam from the CERN SPS with beryllium targets at zero production angle in spills of 4.8\,s duration.
For the test-beam, the momentum of the H6 hadron beam was set to 75\,GeV/$c$, matching the NA62 beam momentum, and \CH\ was placed 440\,m downstream of the H6 target, compared to 70\,m at NA62.
The H6 beam composition at the \CH\ position was estimated to be 4\% kaons, 25\% protons and 71\% pions, similar to the NA62 beam.
The angular divergence of the H6 beam at the \CH\ position was 80\,$\mu$rad both horizontally and vertically, comparable to the 70\,$\mu$rad at NA62 and within the operating tolerance. 
The typical number of particles per spill was $2.8\times10^{5}$.

For the test-beam, \CH\ was equipped with eight ET 9820QB PMTs, operated at a voltage of 2\,kV with a 30\,mV discriminator threshold to achieve single-photoelectron sensitivity.
A pair of scintillator counters, one upstream and one downstream of \CH, provided a trigger signal for each beam particle.
The number of triggers, the number of signals in each PMT, and the numbers of 6-fold, 7-fold and 8-fold coincidences in the PMTs were recorded for each spill.

With the \HH\ pressure at 4.0\,bar and the diaphragm aperture set to 19\,mm, Cherenkov light from pions, kaons and protons passed through the diaphragm. 
In these conditions, the efficiency of each PMT could be measured by the fraction of triggers in which light is detected by the PMT.
The mean PMT efficiency was found to be above 98\%, and the lowest was 95\%.

With the \HH\ pressure set to 3.7\,bar, the pion ring had a radius of 100\,mm, matching the central radius of the diaphragm aperture. With the aperture set to 1.3\,mm, only light from the pion passed through the diaphragm.
The motorised stages were used to align \CH\ parallel to the beam axis, with optimum alignment defined when the PMTs on the top and bottom, and those on the left and right, detected light in the same fraction of triggers.
At the end of the alignment procedure, all the PMTs detected light in 70--72\% of triggers (consistent with the 71\% pion fraction in the beam), which indicated good alignment in both the X and Y directions.
The 1.3\,mm aperture is the smallest, and therefore most sensitive to the alignment, that was used during the test-beam.

\begin{figure}[t]
\centering
\includegraphics[width=0.49\textwidth]{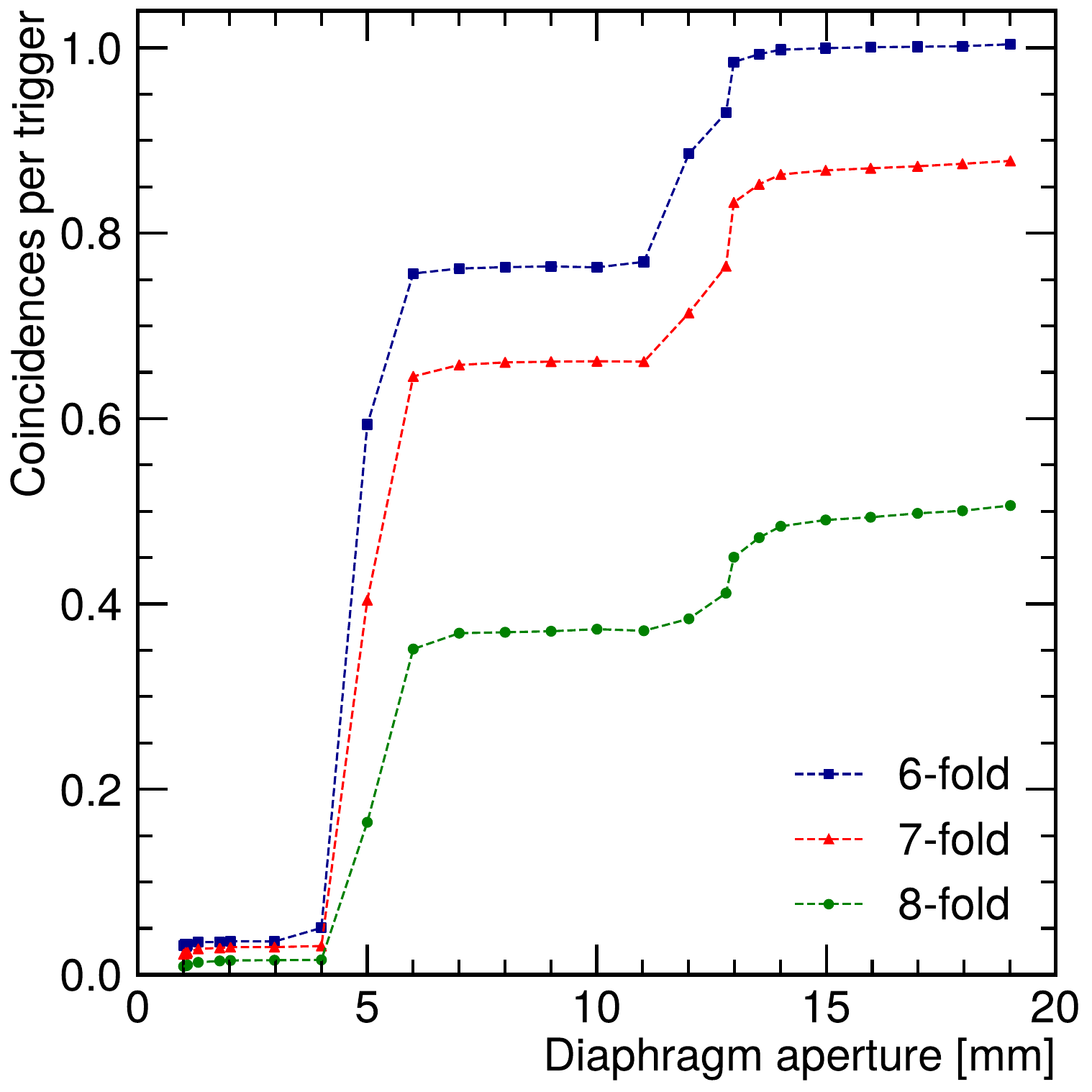}
\caption{Numbers of 6-fold, 7-fold and 8-fold coincidences per trigger as functions of diaphragm aperture at 3.85\,bar, at the H6 test-beam.}
\label{fig:diascan_test}
\end{figure}

The particle identification efficiency was assessed by measuring the number of 6-fold coincidences per trigger while varying the diaphragm aperture from 0.5\,mm to 19.0\,mm at 3.85\,bar.
With the aperture less than 3\,mm, the observed 6-fold coincidences were only due to light from the kaon (figure~\ref{fig:diascan_test}).
Sharp rises at 4\,mm and 11\,mm were observed due to light from the pion and proton passing through the diaphragm, respectively.
At a diaphragm aperture of 19\,mm, a 6-fold coincidence was observed in 99\% of triggers.

\begin{figure}[t]
    \centering
    \subfloat{{\includegraphics[width=0.49\textwidth]{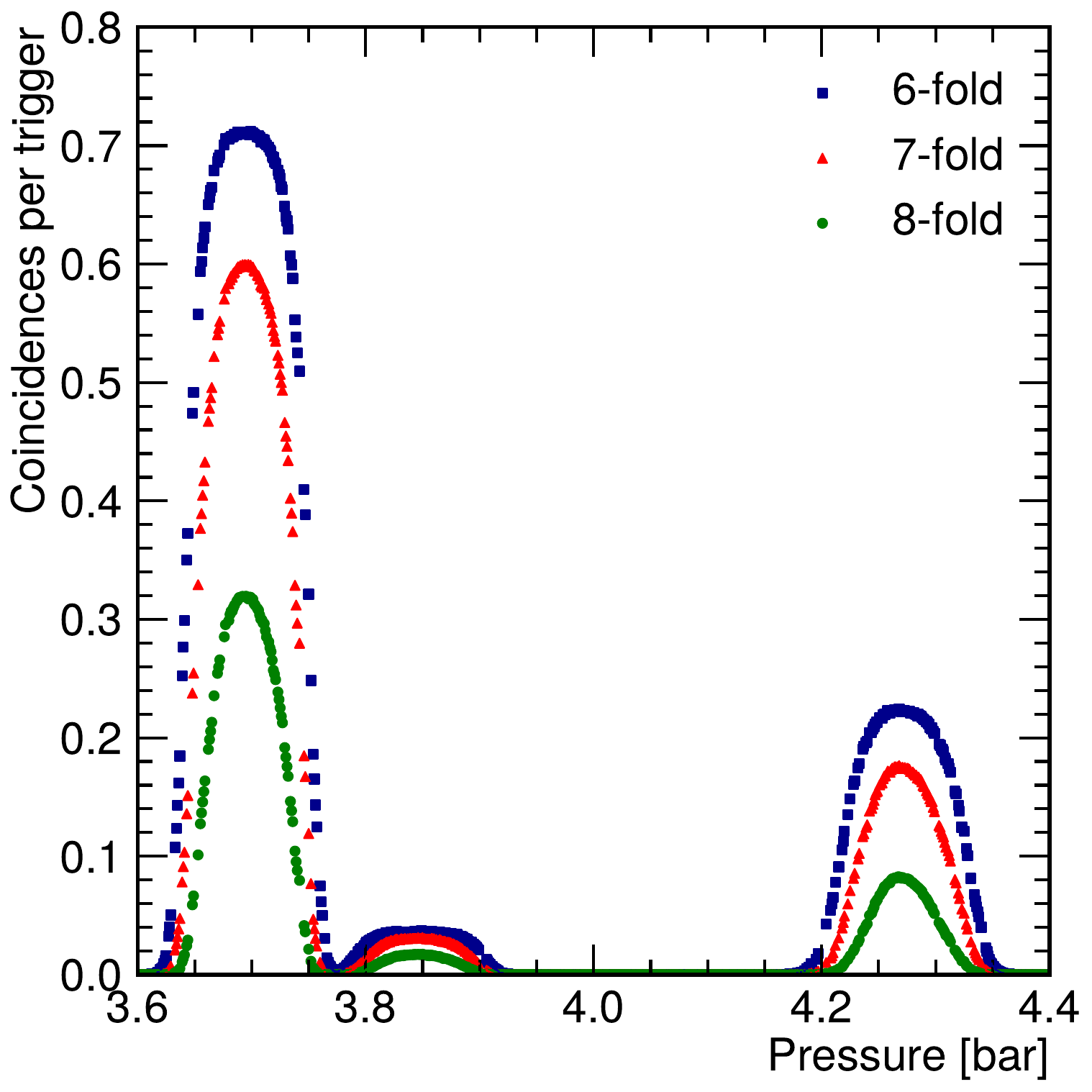} }}%
    \subfloat{{\includegraphics[width=0.49\textwidth]{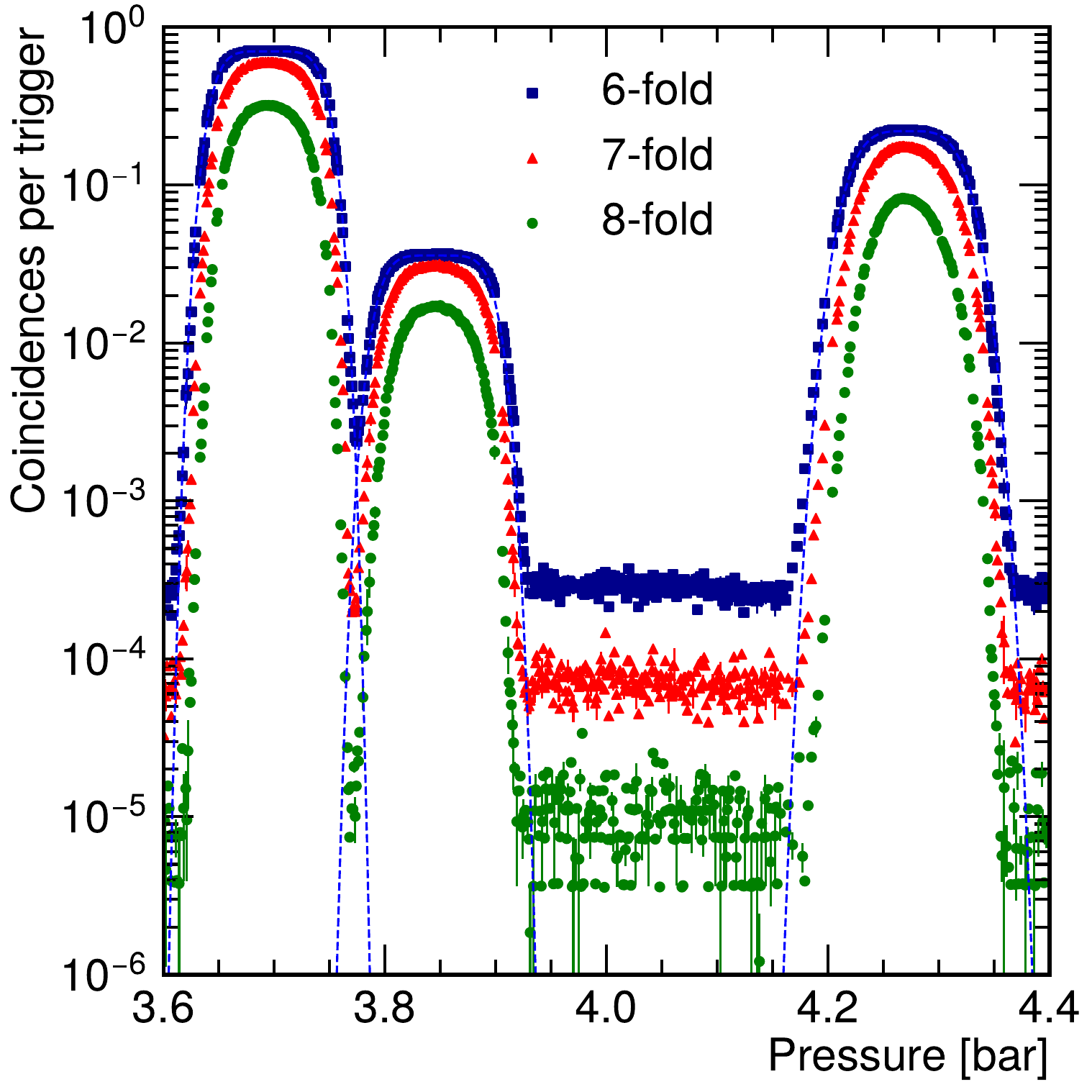} }}%
    \caption{Numbers of 6-fold, 7-fold and 8-fold coincidences per trigger as functions of pressure, at a diaphragm aperture of 1.7\,mm, at the H6 test-beam. The three peaks are from the pion (lowest pressure), kaon and proton (highest pressure), shown in linear scale (left panel) and log scale (right panel). The right panel includes fits to the three peaks (dashed lines).}%
    \label{fig:press17}
\end{figure}

The \CH\ performance with diaphragm apertures ranging from 1.3\,mm to 2.3\,mm was studied by measuring the number of 6, 7 and 8-fold coincidences per trigger while reducing the gas pressure from 4.4 to 3.6\,bar at each aperture setting.
Cherenkov light from the pion, kaon and proton passed through the diaphragm at different pressures, and yielded three distinct peaks, with the centre of the kaon peak at 3.85\,bar.
The optimal diaphragm aperture was found to be 1.7\,mm (figure~\ref{fig:press17}), which maximised the light yield while maintaining an acceptable kaon--pion separation.
The light yield was computed from the number of 6, 7 and 8-fold coincidences per trigger measured at 3.85\,bar:
having defined $\eta_{\mathrm{N}}$ as the probability of an N-fold coincidence for a given trigger, the mean light yield per beam particle was computed in two ways, as 
\[
\lambda= \ln \left[ 1+\frac{8}{\eta_{7}/\eta_{8}-1} \right]
\]
and
\[
\lambda= \ln \left[ 1 + \frac{14}{ \sqrt{4 - 7(1-\eta_{6}/\eta_{8})}-2} \right],
\]
where the two versions were used to cross-check the result. 
The above expressions were obtained using binomial statistics and the Poisson distribution of the number of photoelectrons detected in a single sector~\cite{BOVET}.
The mean light yield per beam particle was found to be 19.1\,photoelectrons, with the two equations giving consistent results.
The kaon--pion separation was determined to be greater than $10^{4}$ by fitting the observed distribution of the fraction of 6-fold coincidences per trigger as a function of pressure in the region of the pion peak, and extrapolating the fitted function to the centre of the kaon peak.
The observed fractions of pions and kaons in the beam, 71\% and 4\% respectively, were in good agreement with the expected beam composition.

\section{\CH\ installation at NA62}

\begin{figure}[t]
    \centering
    \subfloat{{\includegraphics[height=8cm, width=0.68\textwidth]{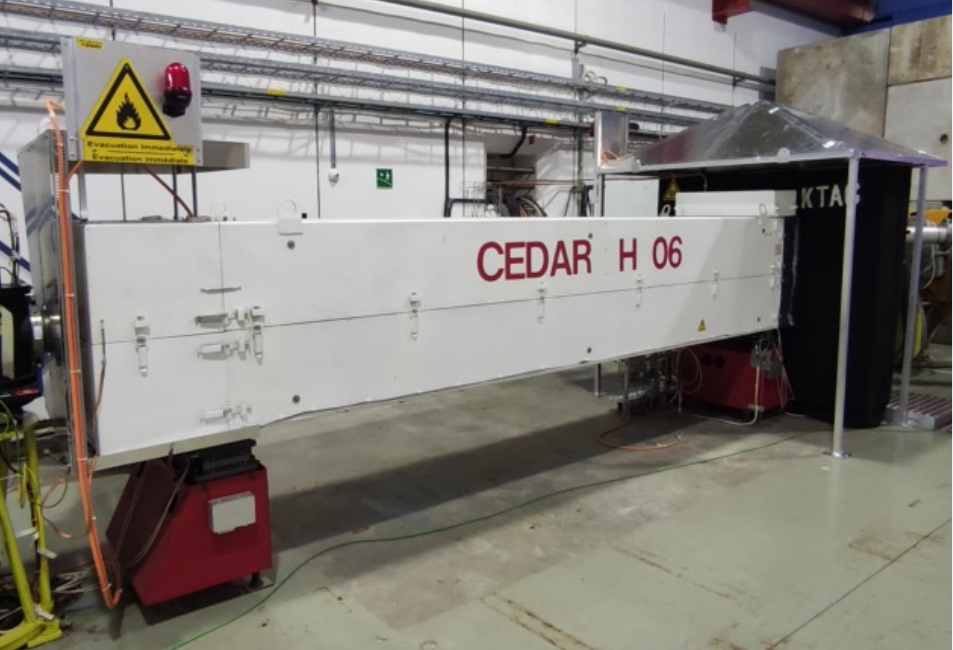} }}%
    \subfloat{{\includegraphics[height=8cm, width=0.30\textwidth]{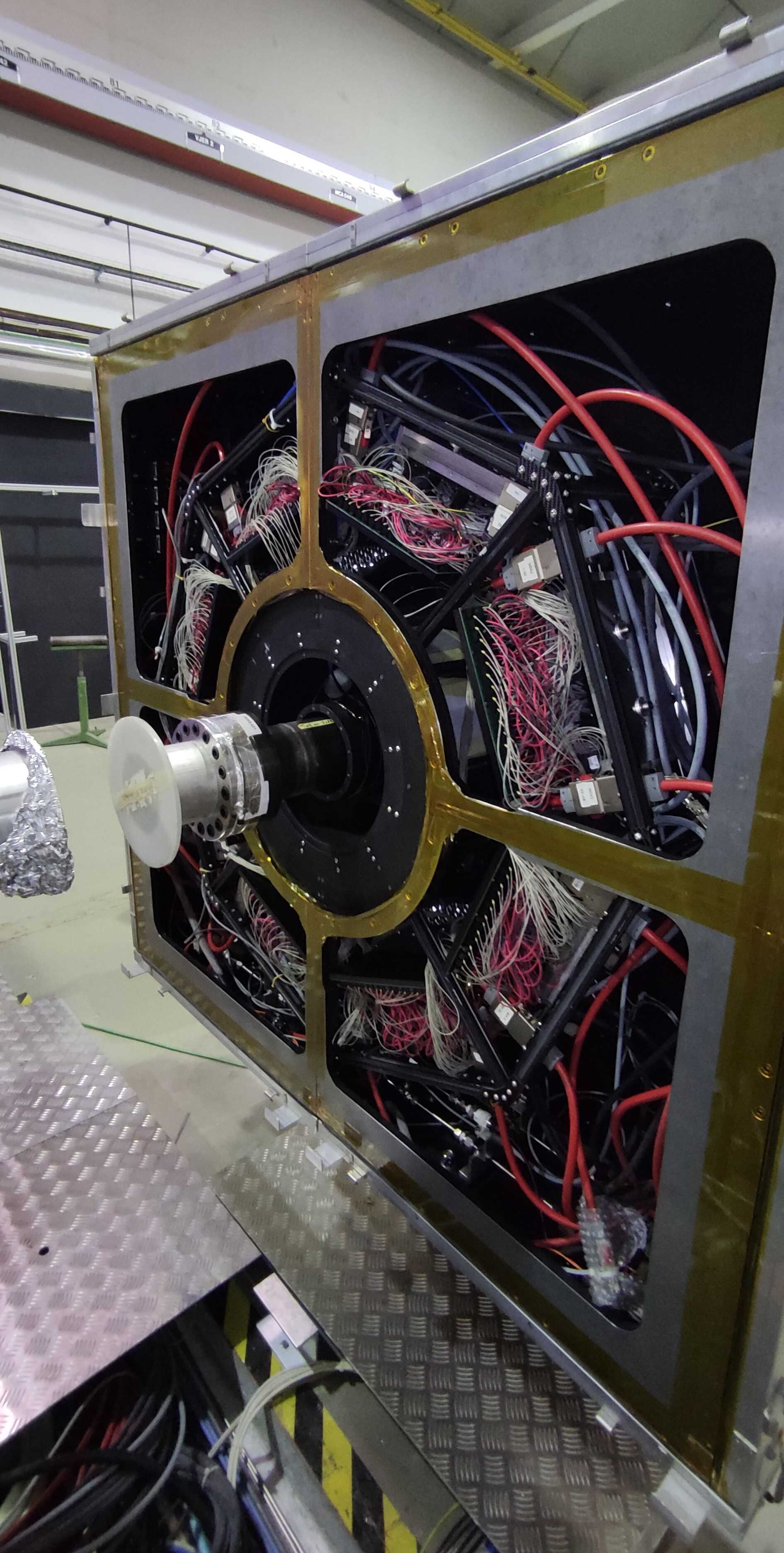} }}%
    \caption{\CH\ (left panel) and KTAG photon-detection system (right panel) during installation in the experimental hall. In the left panel, the downstream end of \CH\ is on the left, and the black enclosure of the KTAG photon-detection system is on the right. In the right panel, the upstream-side of the enclosure is removed, revealing the eight sectors within.}%
    \label{fig:cedarpics}%
\end{figure}

\noindent \CH\ was installed on the NA62 beamline with the KTAG photon-detection system attached in March 2023 (figure~\ref{fig:cedarpics}).
Several modifications were made to the setup to fulfil the safety requirements imposed by the use of \HH\ in the experimental hall,
as the CEDAR is not leak-tight and hydrogen gas is flammable when mixed with air in volume concentrations between 4\% and 75\%.

The volume surrounding any potential \HH\ leak is designated as an explosive atmosphere (ATEX) zone 2.
To avoid sources of ignition, new temperature sensors with suitable certification for use in an explosive atmosphere are used. There are no electrical connectors in the ATEX zones, and all metal parts in the CEDAR-H area are grounded.
Any hydrogen that leaks into confined spaces, such as the KTAG photon-detection system enclosure, is diluted and removed by a flow of nitrogen. Special chimneys have been added to the upstream end of CEDAR-H to avoid accumulation of hydrogen leaking from the gas connections.
A safety interlock protects the experiment in case of a hydrogen leak, identified in the gas control system via an unexpected change in the gas pressure over a defined time period.
The alarm settings are defined by the sensitivity and stability of the pressure gauges.

A dedicated flammable-gas detector, comprising a metal hood and a metal pavilion, monitors \HH\ levels outside of the CEDAR. The metal hood covers the downstream end of \CH, guiding any leaking gas to a hydrogen sensor and physically separating the motorised stages from any potential gas leak.
The metal pavilion covers the upstream end of \CH, plus the KTAG photon-detection system and the gas distribution panel, to guide any leaking gas to another sensor.
The detector issues warning and alarm signals when the \HH\ concentration reaches 0.4\% and 0.8\%, respectively.
In case of an alarm signal, power in the area is cut and the CERN fire and rescue service are automatically informed.

The hydrogen leak rate was evaluated during a 10 day test at 3.85\,bar. The leak rate was found to be 0.2\,litres per day at standard temperature and pressure, satisfying the safety requirements for operating \CH\ given the measured airflow and ventilation in the area.

The rupture of a CEDAR aluminium window would result in an uncontrolled release of \HH\ from the gas vessel.
A safety valve connected to the \CH\ exhaust limits the gas pressure to 5\,bar to avoid stressing the aluminium windows. Each window is tested at 7.5\,bar before installation.
A safety interlock is activated if there is a sudden drop in the gas pressure that indicates one of the aluminium windows has ruptured.

\section{\CH\ commissioning at NA62}

During the \CH\ commissioning, the beam intensity was set to 10\% of the nominal value, which corresponds to a beam particle rate of 60\,MHz.
Dedicated periodic triggers were used to collect $3\times10^{5}$ events per spill.

In standard operating conditions, the KTAG data acquisition records the time and channel ID of each PMT signal in a 100\,ns window defined by the trigger time, with each signal corresponding to a single photoelectron.
The signals are reconstructed into kaon candidates using a clustering algorithm. 
The standard time window for the kaon reconstruction is 2\,ns, however the relatively low kaon rate during the commissioning allowed a 4\,ns time window to be used.
On average, $7\times10^{4}$ kaon candidates were reconstructed in the data collected in each spill.

As the KTAG data acquisition records each photoelectron individually, the definition of the light yield and the kaon tagging requirement differ from those used during the test beam.
The light yield is defined as the mean number of photoelectrons assigned to the kaon candidates. The kaon tagging requires that signals are observed in at least five sectors, as discussed in section~\ref{sec3}.


A coarse angular alignment of \CH\ to the beam is performed with the \HH\ pressure set to the proton peak (4.3\,bar) and the diaphragm aperture set to 6\,mm.
This stage of the alignment is made by balancing the number of signals in the sectors at the top, bottom, left, and right.
The final alignment is performed by maximising the light yield at the pressure of the kaon peak (3.88\,bar) and a diaphragm aperture of 1\,mm.
The optimum alignment yields an average of 21.7 photoelectrons per kaon candidate with the 4\,ns time window.
Changing the alignment in either the vertical or horizontal plane by one increment, corresponding to 7\,$\mu$rad, reduces the number of photoelectrons per kaon candidate by 0.02, showing that the performance is insensitive to small changes in alignment.
The light yield is 15\% larger when using the KTAG PMT arrays instead of the eight PMTs used in the test beam, as expected from the simulation.

\begin{figure}[t]
\centering
\includegraphics[width=0.49\textwidth]{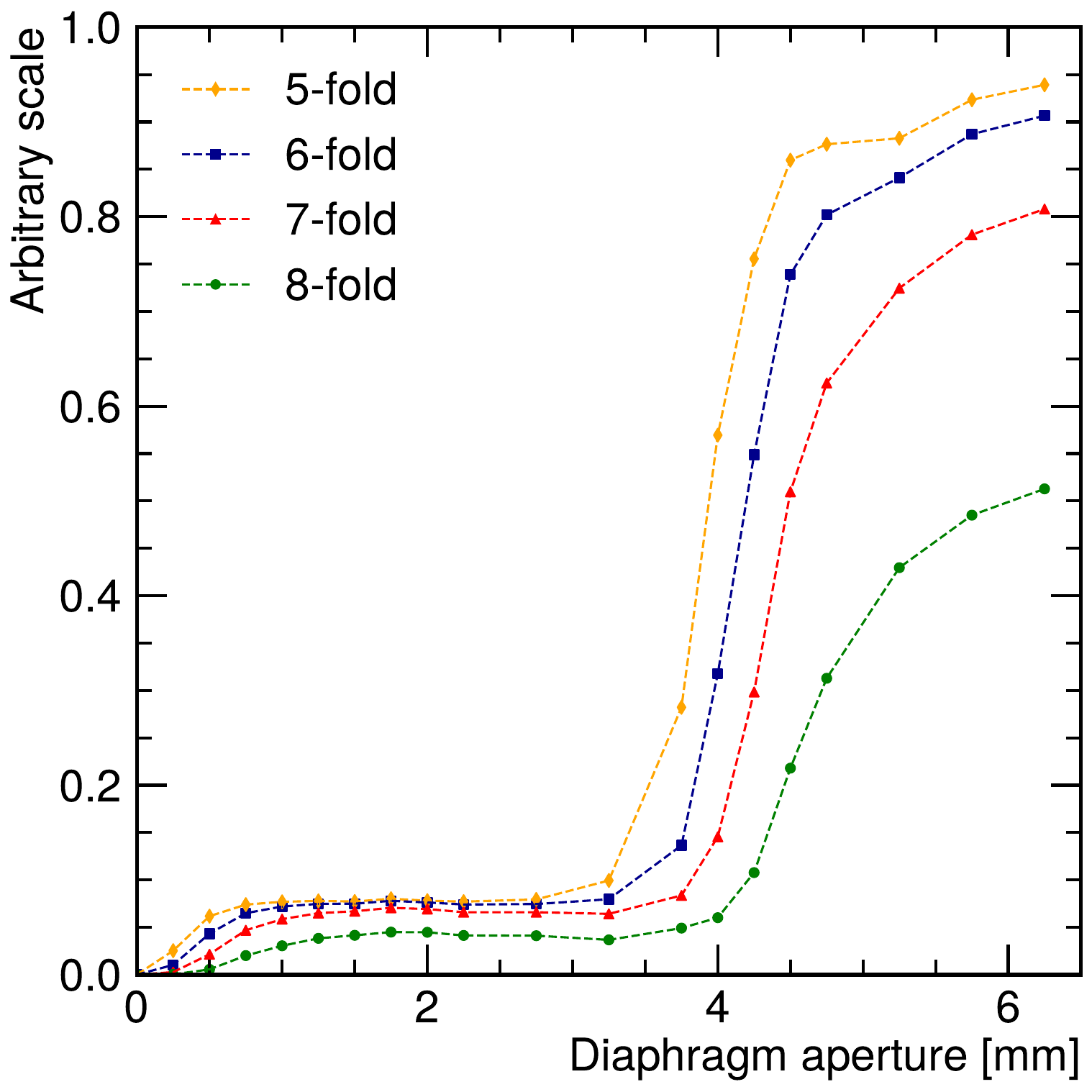}
\caption{Numbers of reconstructed beam particles with 5-fold, 6-fold, 7-fold and 8-fold sector coincidences per beam particle (normalised arbitrarily), as a function of diaphragm aperture at 3.88\,bar, measured using data collected with periodic triggers.}
\label{fig:na62dscan}
\end{figure}

\begin{figure}
    \subfloat{{\includegraphics[width=0.49\textwidth]{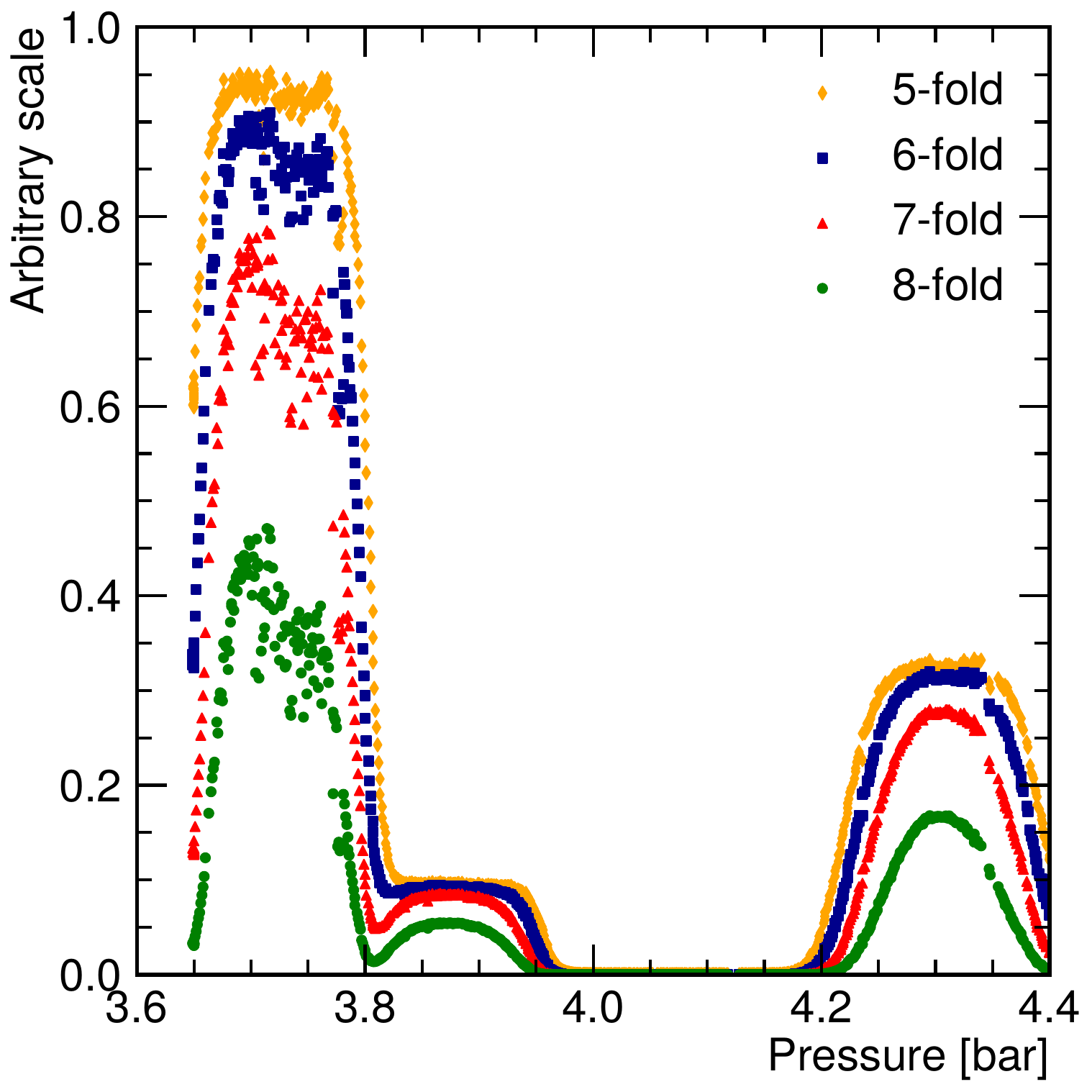} }}%
    \subfloat{{\includegraphics[width=0.49\textwidth]{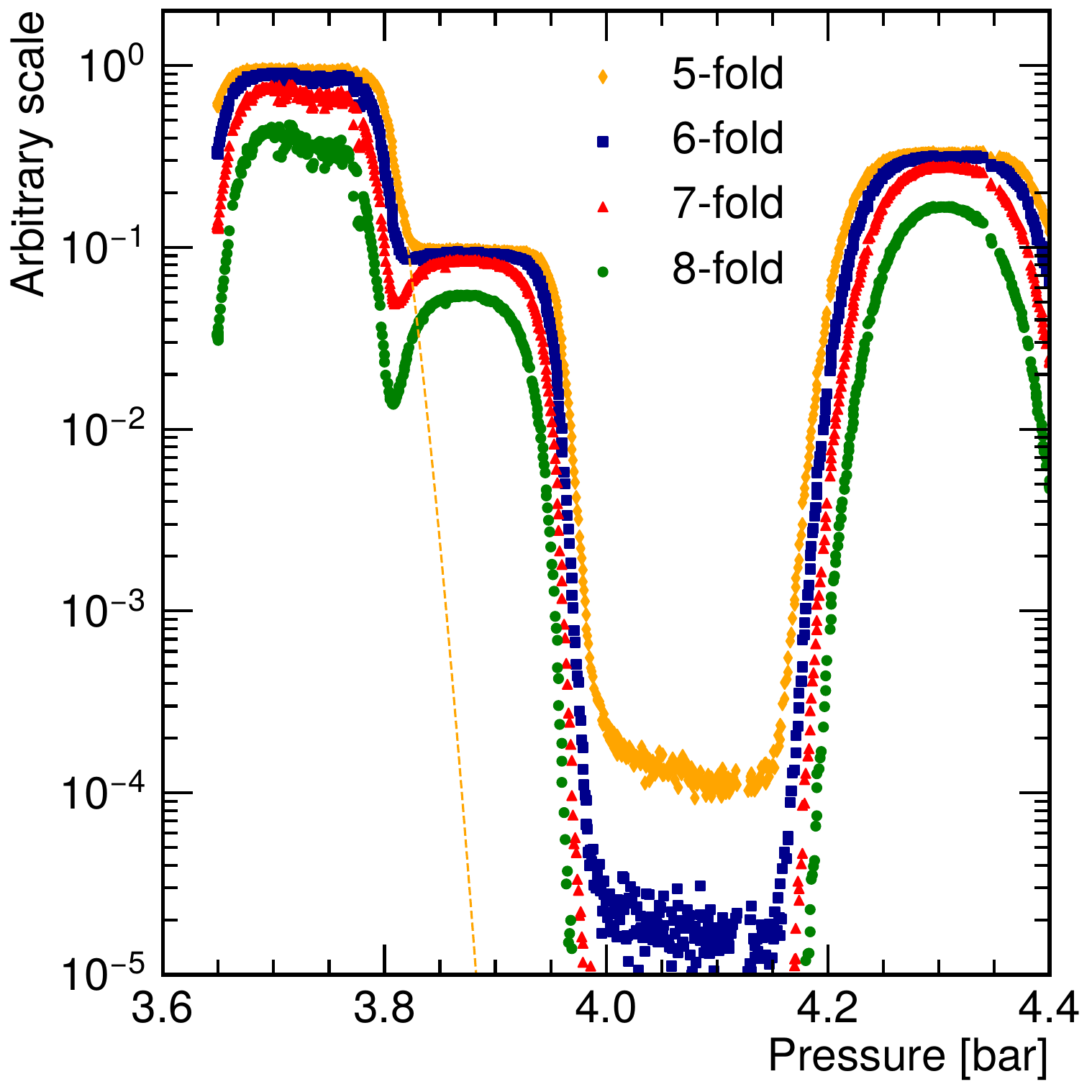} }}%
    \caption{Numbers of reconstructed beam particles with 5-fold, 6-fold, 7-fold and 8-fold sector coincidences in linear scale (left panel) and log scale (right panel). The data were collected with a periodic trigger and a diaphragm aperture of 1.8\,mm, and are normalised to the measured beam intensity. The three peaks correspond to the pion (lowest pressure), kaon and proton (highest pressure). The pion peak is distorted due to limitations of the KTAG data acquisition. The right panel includes a fit to the right side of the pion peak (dashed line).} 
    \label{fig:press17_na62}%

\end{figure}


With the pressure set to the centre of the kaon peak, the optimum aperture is determined by measuring the number of kaon candidates with 5-fold sector coincidences, normalised to the measured beam intensity, at aperture settings ranging from zero to 6\,mm (figure~\ref{fig:na62dscan}). 
The optimum setting is 1.8\,mm, chosen as the smallest value that collects all the light from the kaon.

With the diaphragm aperture set to 1.8\,mm, the numbers of 5, 6, 7, and 8-fold coincidences are measured while reducing the pressure from 4.4 to 3.6\,bar (figure~\ref{fig:press17_na62}).
The centre of the kaon peak is found at 3.88\,bar, and differs from the test-beam value (3.85\,bar) due to $2$\,K higher gas temperature during the commissioning at NA62.
The kaon--pion separation is measured to be better than $10^{4}$ by fitting the distribution of 5-fold sector coincidences as a function of pressure, as described in section~\ref{sec4}. 

\section{\CH\ performance at NA62}

The \CH\ performance in standard NA62 operating conditions is assessed using data collected in 2023, and is compared to that of \CW\ using data collected in 2022.
A sample of charged kaons is obtained by selecting $K^{+}\to\pi^{+}\pi^{+}\pi^{-}$ 
decays fully reconstructed in the STRAW spectrometer.
The decay time is measured with a precision of 200\,ps using the information from the CHOD hodoscopes, and kaon candidates are reconstructed in the KTAG data using the standard 2\,ns time window.

\begin{figure}[t]
\centering
     \subfloat{{\includegraphics[width=0.49\textwidth]{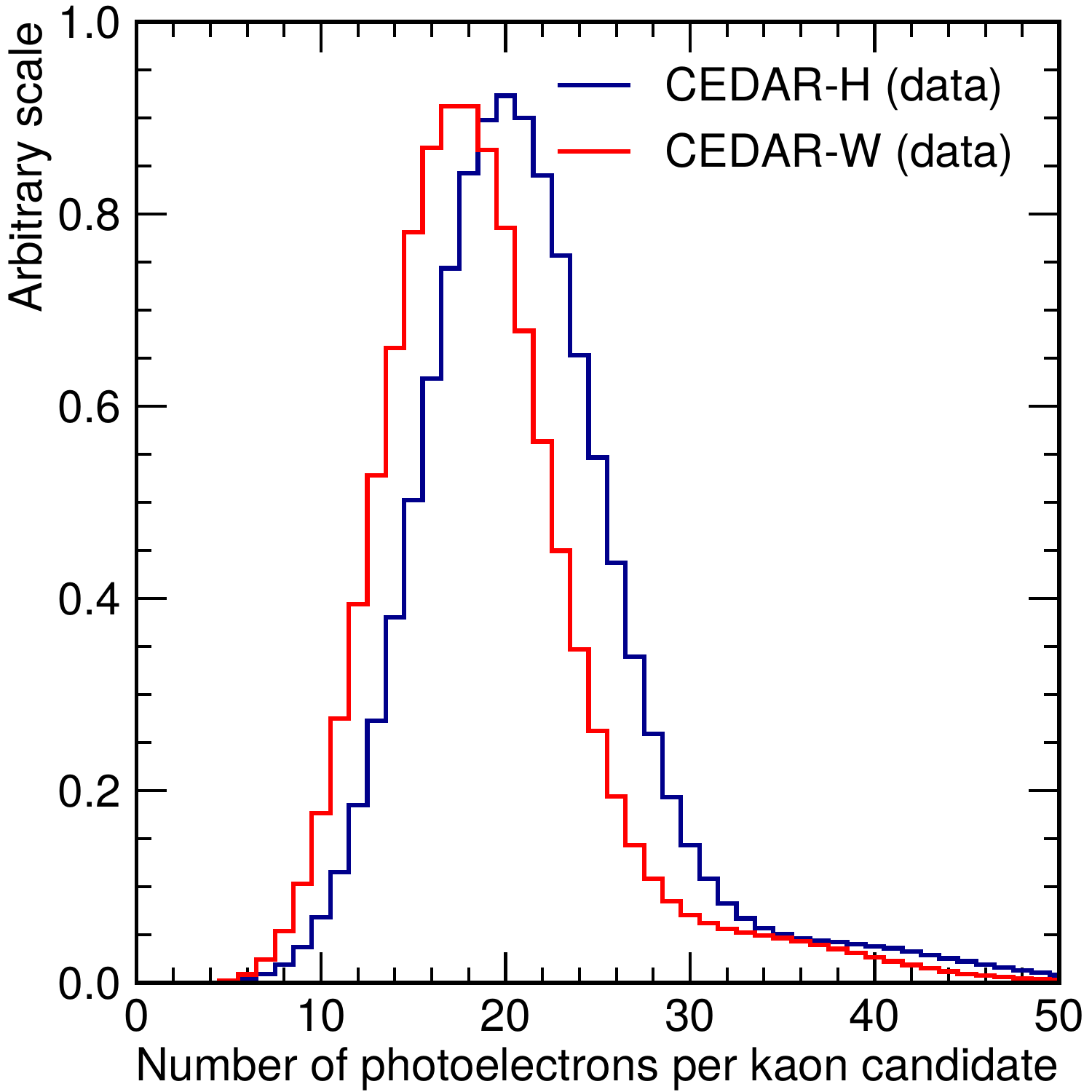} }}%
    \subfloat{{\includegraphics[width=0.49\textwidth]{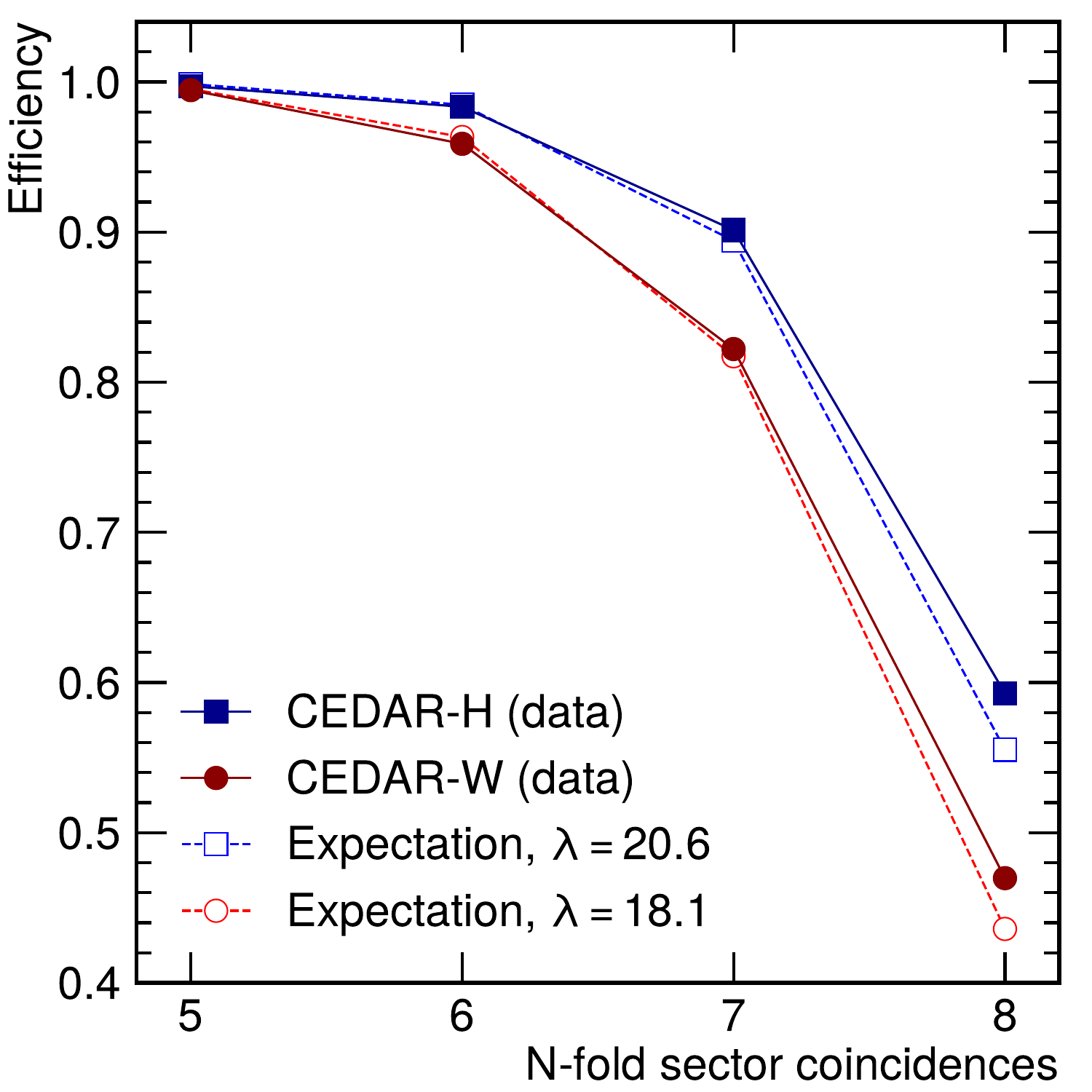} }}%
    \caption{Left: number of photoelectrons per kaon candidate for \CW\ and \CH\ in the data, reconstructed with a 2\,ns time window and normalised to the same integral.
    Right: $K^{+}$ identification efficiency for \CW\ and \CH\ as a function of N-fold sector coincidences, with analytical expectations for the efficiency given the Poisson mean numbers of photoelectrons of 20.6 and 18.1 from fits to the distributions in the left panel.}
    \label{fig:PhotonPlots}%
    
\end{figure}

The light yield and time resolution are measured using kaon candidates with signals in at least 5 sectors and thus satisfy the standard kaon tagging requirement.
The light yield is extracted by fitting a Poisson distribution to the observed number of photoelectrons per kaon candidate between 4 and 24, because a tail towards larger numbers of photoelectrons is observed in the data due to the reconstruction of two coincident $K^{+}$ as a single kaon candidate (figure~\ref{fig:PhotonPlots}, left).
The light yield achieved with \CH\ is 20.6 photoelectrons per kaon candidate, an improvement over 18.1 photoelectrons achieved in 2022 with \CW. 
The \CH\ time resolution is computed to be 66\,ps (compared to 71\,ps for \CW) based on the number of photoelectrons per kaon candidate and the nominal KTAG PMT single-photoelectron time resolution of 300\,ps~\cite{NA62_Det}.

The $K^{+}$ identification efficiency is measured based on the reconstruction of a kaon candidate within 2\,ns of the $K^{+}\to\pi^{+}\pi^{+}\pi^{-}$ decay time that satisfies the tagging requirement.
The $K^{+}$ identification efficiency based on 5-fold coincidences is found to be 99.7\%, compared to 99.5\% for \CW.
The measured efficiency of the two detectors is compared to analytical expectations based on the observed light yield for each detector in figure~\ref{fig:PhotonPlots}, right.
The analytical expectation is computed assuming an ideal detector and a uniform distribution of photoelectrons over the 384 PMTs.
The measured efficiency for 8-fold coincidences is higher than the analytical expectation due to the kaon candidates reconstructed from multiple coincident $K^{+}$.

Reduction of the elastic scattering of beam particles in \CH\ compared to \CW\ is investigated by selecting tracks reconstructed in the STRAW spectrometer with a momentum of 75\,GeV/$c$ that originate upstream of GTK3.
These tracks are selected in data collected via the same minimum-bias hardware trigger in both 2023 and 2022. 
By normalising to the number of $K^{+}\to\pi^{+}\pi^{0}$ decays in each data sample, a 30\% reduction of elastically scattered beam particles originating upstream of GTK3 is observed with \CH\ with respect to \CW, in agreement with simulation.
The corresponding reduction of the trigger rate allowed looser trigger criteria to be imposed in 2023.

\section{Summary}

A kaon tagger for the NA62 experiment has been developed that minimises the amount of material in the path of the beam.
This was accomplished using a North-type CEDAR filled with \HH\ and equipped with a specialised optical system designed using a full simulation of the KTAG.
The detector performance was validated at a test-beam before installation at the experiment. 
Several measures were taken to fulfil safety requirements imposed by the use of \HH.
After the detector was commissioned, data collected in standard operating conditions show that the kaon--pion separation exceeds 
$10^{4}$, the time resolution is 66\,ps, and the $K^{+}$ identification efficiency is 99.7\%.
Each of these values exceeds the kaon tagging requirements.

\section*{Acknowledgements}
We are thankful to V.~Marchand and S.~Mathot for their contributions to
\mbox{CEDAR-H} development and construction; S.~Deschamps, I.~O.~Ruiz and
J.~Tan for setting up the control and DAQ systems for the test-beam;
D.~Banerjee and J.~Bernhard for the beamline, target and controls during
the test-beam; F.~Garnier for the support provided in the KTAG assembly
and disassembly; P.~Boisseaux-Bourgeois, E.~Dho, J.~Gulley, D.~Jaillet,
S.~Marsh, L.~J.~Rowland and H.~Wilkens for expert assistance in assessing
how to reduce health and safety hazards linked to operation with
hydrogen; F.~Corsanego for expert advice on fire prevention measures;
N.~Broca for the integration of the flammable-gas detection system;
O.~O.~Andreassen, N.~El~Kbiri and D.~Lombard for support at the
\mbox{CEDAR-H} conception stage and for the procurement and assembly of
mechanical components; R.~Folch, A.~E.~Rahmoun and M.~B.~Szewczyk for
their assistance and support in the test-beam.

\input{acknowrun2023_optCv2}

\newpage
\section*{The NA62 Collaboration}
\addcontentsline{toc}{section}{\protect\numberline{}The NA62 collaboration}
\input{run2-cedarh-v5}

\end{document}

%% file: acknowrun2023_optCv2.tex
It is a pleasure to express our appreciation to the staff of the CERN laboratory and the technical
staff of the participating laboratories and universities for their efforts in the operation of the
experiment and data processing.

The cost of the experiment and its auxiliary systems was supported by the funding agencies of 
the Collaboration Institutes. We are particularly indebted to: 
F.R.S.-FNRS (Fonds de la Recherche Scientifique - FNRS), under Grants No. 4.4512.10, 1.B.258.20, Belgium;
CECI (Consortium des Equipements de Calcul Intensif), funded by the Fonds de la Recherche Scientifique de Belgique (F.R.S.-FNRS) under Grant No. 2.5020.11 and by the Walloon Region, Belgium;
NSERC (Natural Sciences and Engineering Research Council), funding SAPPJ-2018-0017,  Canada;
MEYS (Ministry of Education, Youth and Sports) funding LM 2023040, Czech Republic;
BMBF (Bundesministerium f\"{u}r Bildung und Forschung) contracts 05H12UM5, 05H15UMCNA and 05H18UMCNA, Germany;
INFN  (Istituto Nazionale di Fisica Nucleare),  Italy;
MIUR (Ministero dell'Istruzione, dell'Universit\`a e della Ricerca),  Italy;
CONACyT  (Consejo Nacional de Ciencia y Tecnolog\'{i}a),  Mexico;
IFA (Institute of Atomic Physics) Romanian 
CERN-RO Nr. 06/03.01.2022
and Nucleus Programme PN 19 06 01 04,  Romania;
MESRS  (Ministry of Education, Science, Research and Sport), Slovakia; 
CERN (European Organization for Nuclear Research), Switzerland; 
STFC (Science and Technology Facilities Council), United Kingdom;
NSF (National Science Foundation) Award Numbers 1506088 and 1806430,  U.S.A.;
ERC (European Research Council)  ``UniversaLepto'' advanced grant 268062, ``KaonLepton'' starting grant 336581, Europe.

Individuals have received support from:
Charles University (Research Center UNCE/SCI/013, grant PRIMUS 23/SCI/025), Czech Republic;
Czech Science Foundation (grant 23-06770S);
Ministero dell'Istruzione, dell'Universit\`a e della Ricerca (MIUR  ``Futuro in ricerca 2012''  grant RBFR12JF2Z, Project GAP), Italy;
the Royal Society  (grants UF100308, UF0758946), United Kingdom;
STFC (Rutherford fellowships ST/J00412X/1, ST/M005798/1), United Kingdom;
ERC (grants 268062,  336581 and  starting grant 802836 ``AxScale'');
EU Horizon 2020 (Marie Sk\l{}odowska-Curie grants 701386, 754496, 842407, 893101, 101023808).

%% file: run2-cedarh-v5.tex
\newcommand{\orcimg}{\raisebox{-0.3\height}{\includegraphics[height=\fontcharht\font`A]{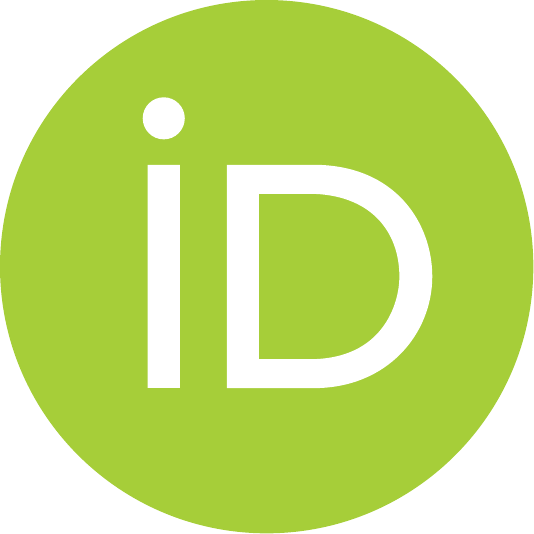}}}
\newcommand{\orcid}[1]{\href{https://orcid.org/#1}{\orcimg}}

%
%

\begin{raggedright}
\noindent
{\bf Universit\'e Catholique de Louvain, Louvain-La-Neuve, Belgium}\\
 A.~Bethani\orcid{0000-0002-8150-7043},
 E.~Cortina Gil\orcid{0000-0001-9627-699X},
 J.~Jerhot$\,${\footnotemark[1]}\orcid{0000-0002-3236-1471},
 N.~Lurkin\orcid{0000-0002-9440-5927}
\vspace{0.5cm}

{\bf TRIUMF, Vancouver, British Columbia, Canada}\\
 T.~Numao\orcid{0000-0001-5232-6190},
 B.~Velghe\orcid{0000-0002-0797-8381},
 V.~W.~S.~Wong\orcid{0000-0001-5975-8164}
\vspace{0.5cm}

{\bf University of British Columbia, Vancouver, British Columbia, Canada}\\
 D.~Bryman$\,${\footnotemark[2]}\orcid{0000-0002-9691-0775}
\vspace{0.5cm}

{\bf Charles University, Prague, Czech Republic}\\
 Z.~Hives\orcid{0000-0002-5025-993X},
 T.~Husek$\,${\footnotemark[3]}\orcid{0000-0002-7208-9150},
 K.~Kampf\orcid{0000-0003-1096-667X},
 M.~Koval\orcid{0000-0002-6027-317X}
\vspace{0.5cm}

{\bf Aix Marseille University, CNRS/IN2P3, CPPM, Marseille, France}\\
 B.~De Martino\orcid{0000-0003-2028-9326},
 M.~Perrin-Terrin\orcid{0000-0002-3568-1956}
\vspace{0.5cm}

{\bf Max-Planck-Institut f\"ur Physik (Werner-Heisenberg-Institut), Garching, Germany}\\
 B.~D\"obrich\orcid{0000-0002-6008-8601},
 S.~Lezki\orcid{0000-0002-6909-774X},
 J.~Schubert\orcid{0000-0002-5782-8816}
\vspace{0.5cm}

{\bf Institut f\"ur Physik and PRISMA Cluster of Excellence, Universit\"at Mainz, Mainz, Germany}\\
 A. T.~Akmete\orcid{0000-0002-5580-5477},
 R.~Aliberti$\,${\footnotemark[4]}\orcid{0000-0003-3500-4012},
 L.~Di Lella\orcid{0000-0003-3697-1098},
 N.~Doble\orcid{0000-0002-0174-5608},
 L.~Peruzzo\orcid{0000-0002-4752-6160}, 
 S.~Schuchmann\orcid{0000-0002-8088-4226},
 H.~Wahl\orcid{0000-0003-0354-2465},
 R.~Wanke\orcid{0000-0002-3636-360X}
\vspace{0.5cm}

{\bf Dipartimento di Fisica e Scienze della Terra dell'Universit\`a e INFN, Sezione di Ferrara, Ferrara, Italy}\\
 P.~Dalpiaz,
 I.~Neri\orcid{0000-0002-9669-1058},
 F.~Petrucci\orcid{0000-0002-7220-6919},
 M.~Soldani\orcid{0000-0003-4902-943X}
\vspace{0.5cm}

{\bf INFN, Sezione di Ferrara, Ferrara, Italy}\\
 L.~Bandiera\orcid{0000-0002-5537-9674},
 A.~Cotta Ramusino\orcid{0000-0003-1727-2478},
 A.~Gianoli\orcid{0000-0002-2456-8667},
 M.~Romagnoni\orcid{0000-0002-2775-6903},
 A.~Sytov\orcid{0000-0001-8789-2440}
\vspace{0.5cm}

{\bf Dipartimento di Fisica e Astronomia dell'Universit\`a e INFN, Sezione di Firenze, Sesto Fiorentino, Italy}\\
 M.~Lenti\orcid{0000-0002-2765-3955},
 P.~Lo Chiatto\orcid{0000-0002-4177-557X},
 I.~Panichi\orcid{0000-0001-7749-7914},
 G.~Ruggiero\orcid{0000-0001-6605-4739}
\vspace{0.5cm}

{\bf INFN, Sezione di Firenze, Sesto Fiorentino, Italy}\\
 A.~Bizzeti$\,${\footnotemark[5]}\orcid{0000-0001-5729-5530},
 F.~Bucci\orcid{0000-0003-1726-3838}
\vspace{0.5cm}

{\bf Laboratori Nazionali di Frascati, Frascati, Italy}\\
 A.~Antonelli\orcid{0000-0001-7671-7890},
 V.~Kozhuharov$\,${\footnotemark[6]}\orcid{0000-0002-0669-7799},
 G.~Lanfranchi\orcid{0000-0002-9467-8001},
 S.~Martellotti\orcid{0000-0002-4363-7816},
 M.~Moulson\orcid{0000-0002-3951-4389}, 
 T.~Spadaro\orcid{0000-0002-7101-2389},
 G.~Tinti\orcid{0000-0003-1364-844X}
\vspace{0.5cm}

\newpage
{\bf Dipartimento di Fisica ``Ettore Pancini'' e INFN, Sezione di Napoli, Napoli, Italy}\\
 F.~Ambrosino\orcid{0000-0001-5577-1820},
 M.~D'Errico\orcid{0000-0001-5326-1106},
 R.~Fiorenza$\,${\footnotemark[7]}\orcid{0000-0003-4965-7073},
 R.~Giordano\orcid{0000-0002-5496-7247},
 P.~Massarotti\orcid{0000-0002-9335-9690}, 
 M.~Mirra\orcid{0000-0002-1190-2961},
 M.~Napolitano\orcid{0000-0003-1074-9552},
 I.~Rosa\orcid{0009-0002-7564-1825},
 G.~Saracino\orcid{0000-0002-0714-5777}
\vspace{0.5cm}

{\bf Dipartimento di Fisica e Geologia dell'Universit\`a e INFN, Sezione di Perugia, Perugia, Italy}\\
 G.~Anzivino\orcid{0000-0002-5967-0952}
\vspace{0.5cm}

{\bf INFN, Sezione di Perugia, Perugia, Italy}\\
 P.~Cenci\orcid{0000-0001-6149-2676},
 V.~Duk\orcid{0000-0001-6440-0087},
 R.~Lollini\orcid{0000-0003-3898-7464},
 P.~Lubrano\orcid{0000-0003-0221-4806},
 M.~Pepe\orcid{0000-0001-5624-4010},
 M.~Piccini\orcid{0000-0001-8659-4409}
\vspace{0.5cm}

{\bf Dipartimento di Fisica dell'Universit\`a e INFN, Sezione di Pisa, Pisa, Italy}\\
 F.~Costantini\orcid{0000-0002-2974-0067},
 M.~Giorgi\orcid{0000-0001-9571-6260},
 S.~Giudici\orcid{0000-0003-3423-7981},
 G.~Lamanna\orcid{0000-0001-7452-8498},
 E.~Lari\orcid{0000-0003-3303-0524}, 
 E.~Pedreschi\orcid{0000-0001-7631-3933},
 J.~Pinzino\orcid{0000-0002-7418-0636},
 M.~Sozzi\orcid{0000-0002-2923-1465}
\vspace{0.5cm}

{\bf INFN, Sezione di Pisa, Pisa, Italy}\\
 R.~Fantechi\orcid{0000-0002-6243-5726},
 F.~Spinella\orcid{0000-0002-9607-7920}
\vspace{0.5cm}

{\bf Scuola Normale Superiore e INFN, Sezione di Pisa, Pisa, Italy}\\
 I.~Mannelli\orcid{0000-0003-0445-7422}
\vspace{0.5cm}

{\bf Dipartimento di Fisica, Sapienza Universit\`a di Roma e INFN, Sezione di Roma I, Roma, Italy}\\
 M.~Raggi\orcid{0000-0002-7448-9481}
\vspace{0.5cm}

{\bf INFN, Sezione di Roma I, Roma, Italy}\\
 A.~Biagioni\orcid{0000-0001-5820-1209},
 P.~Cretaro\orcid{0000-0002-2229-149X},
 O.~Frezza\orcid{0000-0001-8277-1877},
 A.~Lonardo\orcid{0000-0002-5909-6508},
 M.~Turisini\orcid{0000-0002-5422-1891},
 P.~Vicini\orcid{0000-0002-4379-4563}
\vspace{0.5cm}

{\bf INFN, Sezione di Roma Tor Vergata, Roma, Italy}\\
 R.~Ammendola\orcid{0000-0003-4501-3289},
 V.~Bonaiuto$\,${\footnotemark[8]}\orcid{0000-0002-2328-4793},
 A.~Fucci,
 A.~Salamon\orcid{0000-0002-8438-8983},
 F.~Sargeni$\,${\footnotemark[9]}\orcid{0000-0002-0131-236X}
\vspace{0.5cm}

{\bf Dipartimento di Fisica dell'Universit\`a e INFN, Sezione di Torino, Torino, Italy}\\
 R.~Arcidiacono$\,${\footnotemark[10]}\orcid{0000-0001-5904-142X},
 B.~Bloch-Devaux\orcid{0000-0002-2463-1232},
 E.~Menichetti\orcid{0000-0001-7143-8200},
 E.~Migliore\orcid{0000-0002-2271-5192}
\vspace{0.5cm}

{\bf INFN, Sezione di Torino, Torino, Italy}\\
 C.~Biino$\,${\footnotemark[11]}\orcid{0000-0002-1397-7246},
 A.~Filippi\orcid{0000-0003-4715-8748},
 F.~Marchetto\orcid{0000-0002-5623-8494},
 D.~Soldi\orcid{0000-0001-9059-4831}
\vspace{0.5cm}

{\bf Instituto de F\'isica, Universidad Aut\'onoma de San Luis Potos\'i, San Luis Potos\'i, Mexico}\\
 A.~Briano Olvera\orcid{0000-0001-6121-3905},
 J.~Engelfried\orcid{0000-0001-5478-0602},
 N.~Estrada-Tristan$\,${\footnotemark[12]}\orcid{0000-0003-2977-9380},
 R.~Piandani\orcid{0000-0003-2226-8924},
 M.~A.~Reyes Santos$\,${\footnotemark[12]}\orcid{0000-0003-1347-2579},
 K.~A.~Rodriguez Rivera\orcid{0000-0001-5723-9176}
\vspace{0.5cm}

{\bf Horia Hulubei National Institute for R\&D in Physics and Nuclear Engineering, Bucharest-Magurele, Romania}\\
 P.~Boboc\orcid{0000-0001-5532-4887},
 A.~M.~Bragadireanu,
 S.~A.~Ghinescu\orcid{0000-0003-3716-9857},
 O.~E.~Hutanu
\vspace{0.5cm}

{\bf Faculty of Mathematics, Physics and Informatics, Comenius University, Bratislava, Slovakia}\\
 T.~Blazek\orcid{0000-0002-2645-0283},
 V.~Cerny\orcid{0000-0003-1998-3441},
 R.~Volpe$\,${\footnotemark[13]}\orcid{0000-0003-1782-2978}
\vspace{0.5cm}

{\bf CERN, European Organization for Nuclear Research, Geneva, Switzerland}\\
 J.~Bernhard\orcid{0000-0001-9256-971X},
 L.~Bician$\,${\footnotemark[14]}\orcid{0000-0001-9318-0116},
 M.~Boretto\orcid{0000-0001-5012-4480},
 E.~Bravin\orcid{0009-0000-0412-5749},
 F.~Brizioli$\,${\footnotemark[15]}\orcid{0000-0002-2047-441X}, 
 A.~Ceccucci\orcid{0000-0002-9506-866X},
 M.~Ceoletta\orcid{0000-0002-2532-0217},
 M.~Corvino\orcid{0000-0002-2401-412X},
 H.~Danielsson\orcid{0000-0002-1016-5576},
 F.~Duval, 
 L.~Federici\orcid{0000-0002-3401-9522},
 Y.~Fiammingo\orcid{0009-0009-4419-5077},
 E.~Gamberini\orcid{0000-0002-6040-4985},
 A.~Goncalves Martins De Oliveira,
 R.~Guida\orcid{0000-0001-8413-9672}, 
 E.~B.~Holzer\orcid{0000-0003-2622-6844},
 B.~Jenninger,
 Z.~Kucerova\orcid{0000-0001-8906-3902},
 A.~Lafuente Mazuecos\orcid{0009-0009-7230-3792},
 G.~Lehmann Miotto\orcid{0000-0001-9045-7853}, 
 P.~Lichard\orcid{0000-0003-2223-9373},
 K.~Massri\orcid{0000-0001-7533-6295},
 E.~Minucci$\,${\footnotemark[16]}\orcid{0000-0002-3972-6824},
 M.~Noy,
 G.~Rigoletti\orcid{0000-0001-9152-7593}, 
 V.~Ryjov,
 T.~Schneider\orcid{0009-0004-0243-6294},
 J.~Swallow$\,${\footnotemark[17]}\orcid{0000-0002-1521-0911},
 P.~Wertelaers\orcid{0009-0007-4222-7149},
 M.~Zamkovsky\orcid{0000-0002-5067-4789}
\vspace{0.5cm}

{\bf Ecole Polytechnique F\'ed\'erale Lausanne, Lausanne, Switzerland}\\
 X.~Chang\orcid{0000-0002-8792-928X},
 A.~Kleimenova\orcid{0000-0002-9129-4985},
 R.~Marchevski\orcid{0000-0003-3410-0918}
\vspace{0.5cm}

{\bf School of Physics and Astronomy, University of Birmingham, Birmingham, United Kingdom}\\
 J.~R.~Fry\orcid{0000-0002-3680-361X},
 F.~Gonnella\orcid{0000-0003-0885-1654},
 E.~Goudzovski\orcid{0000-0001-9398-4237},
 J.~Hancock\orcid{0000-0003-4808-1551},
 J.~Henshaw\orcid{0000-0001-7059-421X}, 
 C.~Kenworthy\orcid{0009-0002-8815-0048},
 C.~Lazzeroni\orcid{0000-0003-4074-4787},
 C.~Parkinson$\,$\renewcommand{\thefootnote}{\fnsymbol{footnote}}\footnotemark[1]\renewcommand{\thefootnote}{\arabic{footnote}}\orcid{0000-0003-0344-7361},
 A.~Romano\orcid{0000-0003-1779-9122},
 J.~Sanders$\,$\renewcommand{\thefootnote}{\fnsymbol{footnote}}\footnotemark[1]\renewcommand{\thefootnote}{\arabic{footnote}}\orcid{0000-0003-1014-094X}, 
 A.~Sergi$\,${\footnotemark[18]}\orcid{0000-0001-9495-6115},
 A.~Shaikhiev$\,${\footnotemark[19]}\orcid{0000-0003-2921-8743},
 A.~Tomczak\orcid{0000-0001-5635-3567}
\vspace{0.5cm}

{\bf School of Physics, University of Bristol, Bristol, United Kingdom}\\
 H.~Heath\orcid{0000-0001-6576-9740}
\vspace{0.5cm}

{\bf School of Physics and Astronomy, University of Glasgow, Glasgow, United Kingdom}\\
 D.~Britton\orcid{0000-0001-9998-4342},
 A.~Norton\orcid{0000-0001-5959-5879},
 D.~Protopopescu\orcid{0000-0002-8047-6513}
\vspace{0.5cm}

{\bf Physics Department, University of Lancaster, Lancaster, United Kingdom}\\
 J.~B.~Dainton,
 L.~Gatignon\orcid{0000-0001-6439-2945},
 R.~W.~L.~Jones\orcid{0000-0002-6427-3513}
\vspace{0.5cm}

{\bf Physics and Astronomy Department, George Mason University, Fairfax, Virginia, USA}\\
 P.~Cooper,
 D.~Coward$\,${\footnotemark[20]}\orcid{0000-0001-7588-1779},
 P.~Rubin\orcid{0000-0001-6678-4985}
\vspace{0.5cm}

{\bf Authors affiliated with an Institute or an international laboratory covered by a cooperation agreement with CERN}\\
 A.~Baeva,
 D.~Baigarashev$\,${\footnotemark[21]}\orcid{0000-0001-6101-317X},
 D.~Emelyanov,
 T.~Enik\orcid{0000-0002-2761-9730},
 V.~Falaleev$\,${\footnotemark[13]}\orcid{0000-0003-3150-2196}, 
 S.~Fedotov,
 K.~Gorshanov\orcid{0000-0001-7912-5962},
 E.~Gushchin\orcid{0000-0001-8857-1665},
 V.~Kekelidze\orcid{0000-0001-8122-5065},
 D.~Kereibay, 
 S.~Kholodenko$\,${\footnotemark[22]}\orcid{0000-0002-0260-6570},
 A.~Khotyantsev,
 A.~Korotkova,
 Y.~Kudenko\orcid{0000-0003-3204-9426},
 V.~Kurochka, 
 V.~Kurshetsov\orcid{0000-0003-0174-7336},
 L.~Litov$\,${\footnotemark[6]}\orcid{0000-0002-8511-6883},
 D.~Madigozhin\orcid{0000-0001-8524-3455},
 A.~Mefodev,
 M.~Misheva$\,${\footnotemark[23]}, 
 N.~Molokanova,
 V.~Obraztsov\orcid{0000-0002-0994-3641},
 A.~Okhotnikov\orcid{0000-0003-1404-3522},
 I.~Polenkevich,
 Yu.~Potrebenikov\orcid{0000-0003-1437-4129}, 
 A.~Sadovskiy\orcid{0000-0002-4448-6845},
 S.~Shkarovskiy,
 V.~Sugonyaev\orcid{0000-0003-4449-9993},
 O.~Yushchenko\orcid{0000-0003-4236-5115}
\vspace{0.5cm}

\end{raggedright}

%
%
\vspace{0.5cm}
\setcounter{footnote}{0}
\newlength{\basefootnotesep}
\setlength{\basefootnotesep}{\footnotesep}

\renewcommand{\thefootnote}{\fnsymbol{footnote}}
\noindent
$^{\footnotemark[1]}${Corresponding authors: C.~Parkinson, J.~Sanders, email: chris.parkinson@cern.ch, jack.sanders@cern.ch}\\
\renewcommand{\thefootnote}{\arabic{footnote}}
$^{1}${Present address: Max-Planck-Institut f\"ur Physik (Werner-Heisenberg-Institut), Garching, \\
D-85748, Germany} \\
$^{2}${Also at TRIUMF, Vancouver, British Columbia, V6T 2A3, Canada} \\
$^{3}${Also at School of Physics and Astronomy, University of Birmingham, Birmingham, B15 2TT, UK} \\
$^{4}${Present address: Institut f\"ur Kernphysik and Helmholtz Institute Mainz, Universit\"at Mainz, Mainz, D-55099, Germany} \\
$^{5}${Also at Dipartimento di Scienze Fisiche, Informatiche e Matematiche, Universit\`a di Modena e Reggio Emilia, I-41125 Modena, Italy} \\
$^{6}${Also at Faculty of Physics, University of Sofia, BG-1164 Sofia, Bulgaria} \\
$^{7}${Present address: Scuola Superiore Meridionale e INFN, Sezione di Napoli, I-80138 Napoli, Italy} \\
$^{8}${Also at Department of Industrial Engineering, University of Roma Tor Vergata, I-00173 Roma, Italy} \\
$^{9}${Also at Department of Electronic Engineering, University of Roma Tor Vergata, I-00173 Roma, Italy} \\
$^{10}${Also at Universit\`a degli Studi del Piemonte Orientale, I-13100 Vercelli, Italy} \\
$^{11}${Also at Gran Sasso Science Institute, I-67100 L'Aquila,  Italy} \\
$^{12}${Also at Universidad de Guanajuato, 36000 Guanajuato, Mexico} \\
$^{13}${Present address: INFN, Sezione di Perugia, I-06100 Perugia, Italy} \\
$^{14}${Present address: Charles University, 116 36 Prague 1, Czech Republic} \\
$^{15}${Also at INFN, Sezione di Perugia, I-06100 Perugia, Italy} \\
$^{16}${Present address: Syracuse University, Syracuse, NY 13244, USA} \\
$^{17}${Present address: Laboratori Nazionali di Frascati, I-00044 Frascati, Italy} \\
$^{18}${Present address: Dipartimento di Fisica dell'Universit\`a e INFN, Sezione di Genova, \\
I-16146 Genova, Italy} \\
$^{19}${Present address: Physics Department, University of Lancaster, Lancaster, LA1 4YB, UK} \\
$^{20}${Also at SLAC National Accelerator Laboratory, Stanford University, Menlo Park, CA 94025, USA} \\
$^{21}${Also at L.~N.~Gumilyov Eurasian National University, 010000 Nur-Sultan, Kazakhstan} \\
$^{22}${Present address: INFN, Sezione di Pisa, I-56100 Pisa, Italy} \\
$^{23}${Present address: Institute of Nuclear Research and Nuclear Energy of Bulgarian Academy of Science (INRNE-BAS), BG-1784 Sofia, Bulgaria} \\